\documentclass[11pt,a4paper]{article}
\usepackage[utf8]{inputenc}
\usepackage{amsfonts}
\usepackage{amssymb}
\usepackage{mathtools}
\usepackage{graphicx} 
\usepackage{braket}
\usepackage{subcaption}
\usepackage{color}
\usepackage{authblk}
\usepackage{float}
\usepackage[most]{tcolorbox}

\graphicspath{ {pictures/} }

\newcommand{\be}{\begin{equation}}
\newcommand{\ee}{\end{equation}}
\newcommand{\bea}{\begin{eqnarray}}
\newcommand{\eea}{\end{eqnarray}}
\newcommand{\av}[1]{\langle #1 \rangle}
\newcommand{\mink}[1]{\mathbb{M}^#1}
\newcommand{\ds}[1]{\text{dS}_#1}
\newcommand{\flrw}[1]{\text{FLRW}_#1}
\newcommand{\mybox}[1]{\tcbox[sharp corners, boxsep=1mm, boxrule=0.2mm, 
colframe=black,colback=white]{#1}}

\author[1]{\textbf{Joachim Kambor}}
\affil[1]{\textit{Kantonsschule Zug, Fachschaft Physik, Lüssiweg 24, 6302 Zug, Switzerland}}
\author[2]{\textbf{Nomaan X}}
\affil[2]{\textit{Raman Research Institute, Sadashivnagar, Bangalore 560 080, India}}
\title{\textbf{Manifold Properties from Causal Sets using Chains}}
\date{}

\begin{document}
\maketitle
\begin{abstract}
We study the utility of chains defined on causal sets in estimating continuum properties like the curvature, the proper time and the spacetime dimension through a numerical analysis. In particular, we show that in $\ds{2}$ and $\flrw{3}$ spacetimes the formalism of \cite{Roy:2012uz} with slight modifications gives the right continuum properties. We also discuss a possible test of manifoldlikeness using this formalism by considering two models of non-manifoldlike causal sets. This is a part of a broader idea of the geometrical reconstruction of continuum properties given a discrete sub structure, in this case the causal set.          
\end{abstract} 


\section{Introduction}
Causal set quantum gravity is a theory built on the idea that causal ordering is the fundamental building block of Lorentzian geometry \cite{Bombelli:1987aa, Dowker:2005tz, Surya:2019ndm}. This idea is based on theorems proved by Hawking, King, McCarthy and Malament \cite{Hawking:1976fe, Malament:1977fe}. These theorems show that there is a bijection between the conformal class of spacetime metrics and the causal ordering (a partially ordered set). In other words, all geometric information, barring an overall volume factor, about a spacetime manifold can be reconstructed given the causal ordering of events (i.e., points in the spacetime manifold)\footnote{A non-technical discussion of this can be found in Robert Geroch's book \textit{General Relativity from A to B}}.

Subsequently, a lot of work has been done towards constructing topological and geometric properties from the causal ordering. This includes kinematical quantities like dimension estimators obtained with various methods \cite{Myrheim:1978ce,Bombelli:1987aa,Meyer:1988uy,Reid:2003,Roy:2012uz, Glaser:2013pca,Aghili:2018fkd}, topological invariants \cite{Major:2006hv,Major:2009cw}, timelike and spacelike distances \cite{Brightwell:1990ha, Rideout:2009zh,Roy:2012uz,Eichhorn:2019a}, as well as studies of order invariants leading to the curvature components of manifolds that approximate causal sets \cite{Sorkin:2007qi,Sverdlov:2008sx, Benincasa:2010ac,Glaser:2013pca} - for a recent, comprehensive review see \cite{Surya:2019ndm}.
In this article we add to that work by using the abundances of ordered subsets of the causal set, called $k$-chains, for short. We build upon the formalism of \cite{Roy:2012uz}, where the theoretical details were laid out. In particular, we perform a numerical study of this formalism to obtain proper time, spacetime dimension and curvature components from causal sets. The motivation is twofold - first, we want to show that the aforementioned quantities can be robustly determined in simulations of sprinklings into Alexandrov sets of Lorentzian manifolds, thereby establishing explicitly the connection between order invariants of causal sets and geometric properties of the embedding spacetimes. The fact that the formalism of \cite{Roy:2012uz} is directly applicable to non-flat spacetimes is particularly attractive. Second, we study whether the abundances of $k$-chains can be used to identify non-manifoldlike causal sets. 

Although the geometrical reconstruction studied in this paper is related to quantities relevant to the formulation of a dynamical theory, e.g. the scalar curvature, we do not attempt to construct an action. The formalism used here is tied to an expansion in a Riemann normal neighbourhood around some spacetime point, and thus local. The formulation of an action would require gluing together “neighbouring” Alexandrov sets, which is beyond the scope of this work.

A \textit{causal set} $\mathcal{C}$ is a partially ordered set with an order-relation $\prec$ which is $\forall\,x,y,z\in\mathcal{C}$
\begin{enumerate}
    \item \textit{Antisymmetric:} $x\prec y\prec x\Rightarrow x=y$
    \item \textit{Transitive:} $x\prec y\prec z\Rightarrow x\prec z$
    \item \textit{Locally finite:} $|\{z\in\mathcal{C}|x\prec z\prec y\}|<\infty$
\end{enumerate}
Here $|\cdot|$ denotes the cardinality of a set. The elements of $\mathcal{C}$ represent spacetime events\footnote{This is not true in general. In the case of non-manifoldlike causal sets, elements do not represent events.} and hence the order-relation $\prec$ denotes the causal order between the events. If $x\prec y$ we say ``$x$ causally precedes $y$". Causal relations on a Lorentzian manifold (without closed timelike curves) obey the first two conditions. Condition 3 ensures that there are a finite number of events in any causal interval; this brings in discreteness. $\mathcal{C}$ is characterized by a matrix, usually denoted by $C$, which is called the \textit{causal matrix}. This matrix is upper diagonal (by convention) with entries $C_{ij}=1$ when $i\prec j$ and $C_{ij}=0$ otherwise. 

\begin{figure}[!htp] 
\centering
  \includegraphics[width=6cm,keepaspectratio]{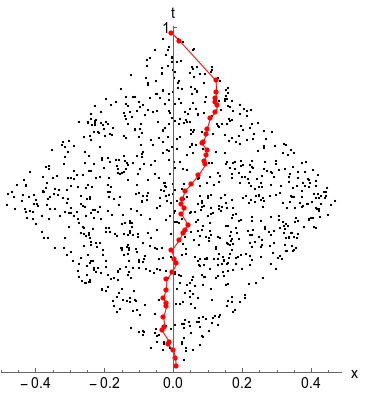}
  \caption{Example of a $k$-chain in a causal set sprinkled in a region of $\mink{2}$.}
  \label{chainsinas}
  \end{figure}

A \textit{$k$-chain} is a totally ordered subset of $\mathcal{C}$. We define $C_k\,(k\geq 1)$ to mean a chain of length $k-1$ which consists of $k$ ordered elements $u_1\prec....\prec u_k$. In Fig (\ref{chainsinas}) we show an example of such a chain for a region of $\mink{2}$.
  
The information about the volume is obtained by requiring that the number of causal set elements $N$ in a spacetime region of volume $V$ is given by
\be
    N=\rho V
    \label{numbervol}
\ee
where $\rho$ is a density factor which also defines a fundamental discreteness scale $\rho^{-1}$ (this maybe the Planck scale for example). Causal set theory also requires\footnote{This is called the fundamental conjecture or the \textit{Hauptvermutung} of causal set theory.} that if $\mathcal{C}$ is approximated by a spacetime $(\mathcal{M},g)$ then $(\mathcal{M},g)$ is unique up to scales $>\rho^{-1}$.

From a mathematical point of view a causal set is a vastly more general object than a spacetime manifold. More precisely, a causal set generated at random\footnote{This can be thought of as generating an upper diagonal matrix $C_{ij}$ with $1$s and $0$s placed randomly, so long as they satisfy the definition of the causal set.} may or may not correspond to a manifold. Even if this were the case, it may correspond to an arbitrary spacetime with any dimension. In order to check if the properties we obtain from the causal set correspond to anything meaningful in the continuum, we must know the corresponding continuum properties in advance. This is not possible if we work with randomly generated causal sets, hence we work with \textit{sprinkled causal sets}. 

To obtain a sprinkled causal set we take a region of a known spacetime and discretize it. The events thus obtained form the causal set and the order relations are inherited from the continuum. To ensure that such a process is covariant i.e. the points picked are not based on any specific coordinate system we use a random Poisson discretization \cite{Bombelli:2006nm}. This implies that the probability of picking $n$ points from a spacetime region of volume $V$, given a fundamental discreteness scale $\rho^{-1}$ is
\be \label{poisson}
    P_V(n)=\frac{(\rho V)^n e^{-\rho V}}{n!}
\ee
which also gives us $\langle n \rangle=N=\rho V$ in a statistical sense. So in working with sprinkled causal sets, we sprinkle into a given spacetime region multiple times and obtain the quantities of interest each time. We work with averages of these quantities\footnote{c.f. Eq. \eqref{avgc}} and their combinations while comparing them with the relevant continuum properties.
\\

In section \ref{curvature estimates} we estimate $R$, $R_{00}$ and the proper time $\tau$ from a given causal set derived from regions in $\ds{2}$ and $\flrw{3}$ spacetimes. We begin with an outline of the theoretical framework based on \cite{Roy:2012uz,Meyer:1988uy} and its generalization. We then discuss the choice of region in each spacetime and derive useful properties of these regions. Finally, we present the results for these cases. While the estimation of $\tau$ is robust, the values of $R$ and $R_{00}$ are prone to higher fluctuations. However, working with the generalized framework and averaging over more sprinklings allows us to keep the fluctuations in check and improve statistics.   

In Section \ref{dimension} we first recall results from $d=2,3,4$ in Minkowski spacetime. In the curved spacetime regions of $\ds{2}$ and $\flrw{3}$ we show that the dimension estimator of \cite{Roy:2012uz}, with appropriate generalizations gives the expected dimension. In Section \ref{non-manifold} we discuss a possible method to distinguish causal sets that are not approximated by manifolds. We apply the formalism to causal sets obtained from a coupled chains model and a Sierpinski mesh. These do not correspond to manifolds and we find that indeed the dimension estimator does not converge. Other quantities also fluctuate wildly. Finally in Section \ref{conclusions} we discuss our results and the broader context for this work.

\section{Curvature Estimates}
\label{curvature estimates}

In a causal set approximated by a region of a Lorentzian manifold, the distribution of k-chains, $C_k$, can be employed to determine local geometrical and topological properties. Meyer \cite{Meyer:1988uy} obtained an analytical expression for averages of $C_k$ over sprinklings in an Alexandrov Set (AS)\footnote{For 2 causally related points $x\prec x'$ an \textit{Alexandrov Set} or \textit{causal diamond} is defined as the region $J^+(x)\cap J^-(x')$.} of $n$-dimensional Minkowski space. The procedure was generalized for manifolds with curvature, using an expansion in a Riemann normal neighbourhood (RNN) \cite{Roy:2012uz} around some spacetime point $x_0$. These results allow us to extract the scalar curvature $R$ as well as the time-component of the Ricci tensor from purely order theoretic information. 
Below we first gather the basic formulae which will be used in our approach. In the two cases that we consider - $\ds{2}$ and $\flrw{3}$, we discuss multiple methods to determine the curvature parameters.

The distribution of $k$-chains in an AS sprinkled in $\mathbb{M}^n$ is\footnote{A modified version of this expression was calculated in \cite{Aghili:2018fkd}. However since the number of sprinkled points we use in this work are large, the deviations from this expression are negligible.}
\be
\av{C_k}_\eta=\rho^k\,\zeta_k\tau^{kn},\quad\quad\zeta_k=\zeta_0^k\chi_k
\label{avgc}
\ee
where the average is taken over sprinklings.
Here $\tau$ is the size of the AS, $\rho$ is the sprinkling density, $\zeta_0=\dfrac{A_{n-2}}{2^{n-1}n\,(n-1)}$ and
\be \label{chikdef}
\chi_k=\dfrac{1}{k}\bigg(\dfrac{\Gamma(n+1)}{2}\bigg)^{k-1}\dfrac{\Gamma(n/2)\,\Gamma(n)}{\Gamma(kn/2)\,\Gamma((k+1)n/2)}
\ee
Putting in $A_{n-2}=\dfrac{2\pi^{(n-1)/2}}{\Gamma((n-1)/2)}$ (the "surface area" of the unit sphere $S^{n-2}$), we get
\be
\zeta_0=\dfrac{\pi^{(n-1)/2}}{2^{n-2}n\,(n-1)}\dfrac{1}{\Gamma((n-1)/2)}
\ee

In a spacetime with curvature $\left(\mathcal{M},g\right)$, the average over sprinklings of the distribution of $k$-chains in an AS may be obtained by an expansion in an RNN. To leading order in an expansion in the proper time span $\tau$ of the AS it is \cite{Roy:2012uz}
\be
\av{C_k}=\av{C_k}_\eta[1+\alpha_kR(0)\tau^2+\beta_kR_{00}(0)\tau^2]+\mathcal{O}(\tau^{kn+3})
\label{avgcexpansion}
\ee
where
\be
\alpha_k=\frac{-nk}{12(nk+2)(n(k+1)+2)},\quad\quad\beta_k=\frac{nk}{12(n(k+1)+2)}
\ee

Once the distributions of $k$-chains in some AS of $\left(\mathcal{M},g\right)$ are known, the expansion (\ref{avgcexpansion}) may be used to extract $\tau$, $R$ and $R_{00}$ of this local neighbourhood of the manifold\footnote{An alternative way of determining $\tau$ uses the length of the longest chain \cite{Brightwell:1990ha, Rideout:2009zh}.}. The explicit formulae given in \cite{Roy:2012uz} can be generalized as follows. Define 
\bea
Q_{k,\lambda}&\equiv&\bigg(\frac{\av{C_k}}{\rho^k\zeta_k}\bigg)^{\lambda/k}=\frac{1}{\zeta_0^\lambda}\bigg(\frac{\av{C_k}}{\rho^k\chi_k}\bigg)^{\lambda/k} \label{Qklam}\\
K_{k,\lambda}&\equiv&((k+1)n+2)\,Q_{k,\lambda} \label{Kklam}\\
J_{k,\lambda}&\equiv&(kn+2)\,K_{k,\lambda} \label{Jklam}
\eea
Then
\bea
R(0)&=&\frac{-6\,(k_1n+2)((k_1+1)n+2)((k_1+2)n+2)}{\lambda\,n^3\tau^{\lambda n+2}}\times\label{Rgen}\nonumber\\
&&\quad\quad \left(K_{k_1,\lambda}-2K_{k_1+1,\lambda}+K_{k_1+2,\lambda}\right)
\eea
\begin{align} \label{R00gen}
&\quad\quad\quad R_{00}(0)=\dfrac{-12\,((k_1+1)n+2)((k_1+2)n+2)}{\lambda\,n^3\tau^{\lambda n+2}}\times\\
&\left((k_1n+2)Q_{k_1,\lambda}-2((k_1+3/2)n+2)Q_{k1+1,\lambda}+((k_1+3)n+2)Q_{k_1+2,\lambda}\right)\nonumber
\end{align}
\be \label{Tgen}
\tau^{\lambda n}=\frac{1}{2n^2}\left(J_{k_1,\lambda}-2J_{k_1+1,\lambda}+J_{k_1+2,\lambda}\right) 
\ee

Compared to the expressions given in \cite{Roy:2012uz} two changes have been made: (1) in the definition of $Q_{k,\lambda}$ the power 3 is replaced by a real, positive parameter $\lambda$ and (2) instead of using the first three moments $Q_{k,\lambda}$, $k=1,2,3$, we may use more general consecutive $k$-values $k_1,\ k_1+1,\ k_1+2$, $k_1\geq 1$.
For further use we define a few more quantities of interest
\bea
\overline{\av{C_k}}\equiv\frac{\av{C_k}}{\av{C_1}^k},\qquad
\overline{Q_{k,\lambda}}\equiv\frac{Q_{k,\lambda}}{\tau^{\lambda n}}\\
\text{and}\quad\quad
\Delta\,\overline{\av{C_k}}_\text{norm}\equiv\Bigl(\frac{\overline{\av{C_k}}}{\chi_k(n)}\Bigr)^{1/k}-1
\label{delCkbarnorm}
\eea
In particular, the last quantity gives a normalized measure of deviation from the Minkowski case (where it is 0). To lowest order in the expansion \eqref{avgcexpansion} it reads
\be
\Delta\,\overline{\av{C_k}}_\text{norm}^\text{RNN}=
\Bigl(\frac{1}{k}\alpha_k-\alpha_1\Bigr) R(0)\tau^2+\Bigl(\frac{1}{k}\beta_k-\beta_1\Bigr) R_{00}(0)\tau^2+\mathcal{O}(\tau^3)
\label{delCkbarnormRNN}
\ee

Before proceeding to the simulations we would like to mention two points relevant to the application of the formalism. First, the individual determination of proper the time $\tau$ and the curvature parameters $R, R_{00}$ requires the introduction of a scale. In the formalism, this scale is hidden in the definition of the quantities $Q_{k,\lambda}$, which depend on the density $\rho$. If the underlying manifold is known, it is natural to introduce a scale via the proper time span of the region to be sprinkled. The determination of $\tau$ according to Eq. \eqref{Tgen} is then to be understood as a consistency check, and the curvature parameters can be extracted from the scale independent combinations $R\,\tau^2$ and $R_{00}\,\tau^2$. We will follow this procedure in section \ref{curvature estimates} where we sprinkle into known manifolds. If, on the other hand, only the causal relations are known and thus no scale is available, we are restricted to the above mentioned scale independent quantities. This more general setting applies to section \ref{non-manifold}, where we consider more abstract causal sets. A second restriction is due to the perturbative nature of the approach. The expansion parameters of the underlying Eq.(\ref{avgcexpansion}) are the dimensionless quantities $R\,\tau^2$ and $R_{00}\,\tau^2$. The RNN and accordingly the proper time span $\tau$ must be chosen small enough in order for the expansion to converge rapidly. We note that the coefficients $\alpha_k,\ \beta_k$ appearing in the leading order corrections are small, thus further suppressing the curvature corrections. $\alpha_k$ decreases with $k$, whereas $\beta_k$ saturates. The large $k$ limit of these quantities is
\be
\lim_{k\to\infty}(\alpha_k,\beta_k) \to \bigg(\frac{-1}{12 k n},\frac{1}{12}\bigg)
\ee
Therefore we require that
\bea
|R\,\tau^2|&<<&\frac{1}{\alpha_{k,\text{min}}}=12\,n\\
|R_{00}\,\tau^2|&<<&\frac{1}{\beta_{k,\text{min}}}=12
\eea

In the simulations shown below we chose $\tau$ such that the leading order corrections to $\overline{\av{C_k}}$ are at most 10 \%. The corrections to $\overline{Q_{k,\lambda}}$ are larger, in particular for smaller $k$ values.

\subsection{de Sitter Spacetime}
\label{curvatureds}

\begin{figure}[h!]
 \begin{subfigure}[b]{0.5\textwidth}
 \includegraphics[height=6cm]{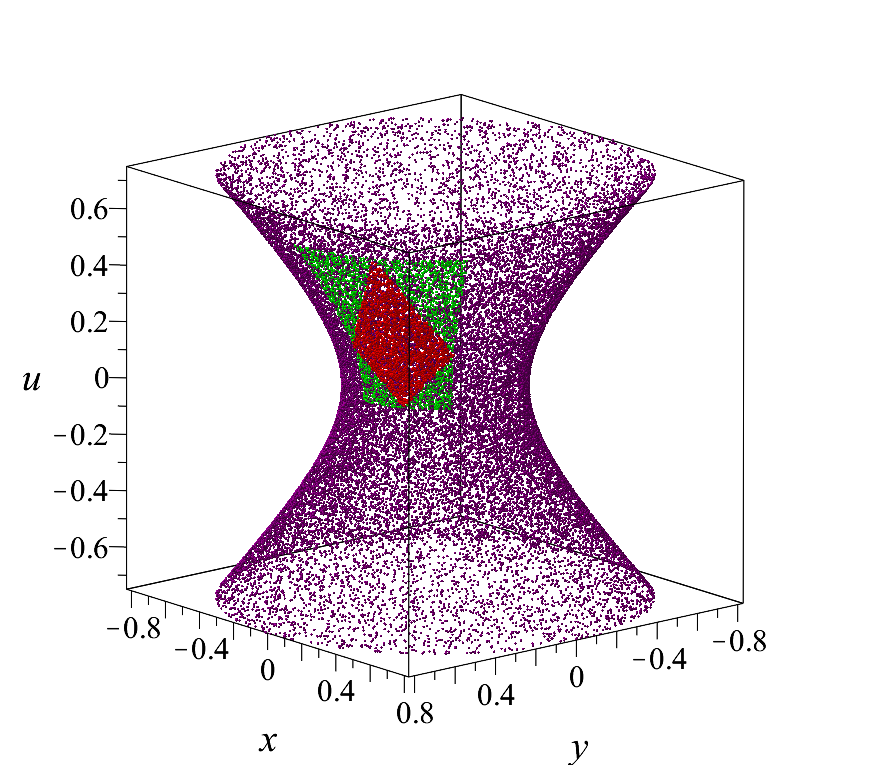}
  \caption{}
  \label{deS_Hyperboloid}
  \end{subfigure}
  \begin{subfigure}[b]{0.5\textwidth}
 \includegraphics[width=5cm]{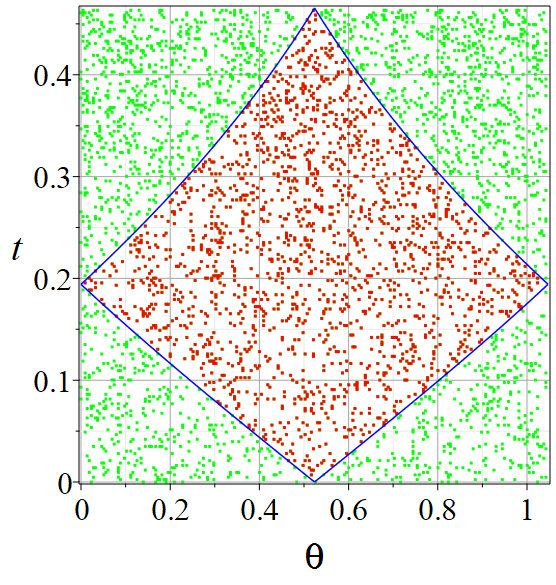}
  \caption{}
 \label{deS_AS_boundaries}
  \end{subfigure}
   \caption{(a) shows a causal set sprinkled in $\ds{2}$ spacetime, a corresponding AS is shown in red and the region of interest in green. (b) shows the shape of null lines.}
\end{figure}

In this example we work with $n=2$. $\ds{2}$ can be thought of as a rotational hyperboloid embedded in three dimensional Minkowski space $\mink{3}$. Denoting coordinates $u, x, y$ in this 3-dimensional space, the hyperboloid is defined by the condition
\be
-u^2+x^2+y^2=\alpha^2 \label{defhyperbol}
\ee
A visualization of this construction is shown in Figure \ref{deS_Hyperboloid}.

The curves $u=\text{const.}$ are circles with radii $r\equiv \sqrt{x^2+y^2}=\sqrt{\alpha^2+u^2}$. The parameter $\alpha$ thus has the meaning of the smallest possible radius at the waist of the hyperboloid.  

\subsubsection{Choice of region, coordinates and parameters}

The invariant line element of the embedding space is given by 
\be
ds^2=-du^2+dx^2+dy^2
\ee
The metric on the hyperboloid is the induced metric, restricted to the surface \eqref{defhyperbol}.

$\ds{2}$ can be described with the two coordinates $t, \phi$. $\phi$ is the angle around the $u$-axis and $t$ is implicitly given by
\be
u=\alpha\ \sinh\left(\frac{t}{\alpha}\right)
\ee
The scalar curvature is given by
\be
R=\frac{2}{\alpha^2}
\ee
It is positive and constant on the entire spacetime. 

In the simulations we need to sprinkle into a region of $\ds{2}$ which is well suited to construct an AS between two points $P_1, P_2$. The region can be parameterized as follows:
\bea
0 &\leq& t \leq t_0 \\
0 &\leq& \phi < \phi_0
\eea
The volume of this region is
\be
V_\text{region}=\int_{0}^{\phi_0}d\phi\;\int_{0}^{t_0}dt\sqrt{-\det\left(g\right)}=\phi_0\ \alpha^2\sinh\left(\frac{t_0}{\alpha}\right)
\label{Aregion}
\ee

The range of coordinates has to be chosen such that all points belonging to the AS are contained in the region. 
A further restriction on the choice of the parameters arises from the expansion in an RNN,
$t_0$ must be chosen such that the resulting region is small enough for the $\tau^2$-expansion to be valid.

\subsubsection{Creating the Alexandrov Set}

The boundaries of the AS, i.e. the light rays in $\left(t,\phi\right)$ - space are given by curves as shown in Fig. \ref{deS_AS_boundaries}.

These light rays are defined by the condition 
\bea
ds^2&=&dt^2-\alpha^2\: \cosh^2\left(\frac{t}{\alpha}\right)\: d\phi^2=0\\
\ \frac{dt}{d\phi}&=&\pm\: \alpha\cosh\left(\frac{t}{\alpha}\right) \label{DGL_lightrays}
\eea
The plus sign in Eq. \eqref{DGL_lightrays} corresponds to the ray going out from $P_1$ to the right, the minus sign to the ray coming into $P_2$ from the right. Integrating and using appropriate boundary conditions we obtain
\bea
t_+\left(\phi\right)&=&\alpha\:\ln\left[ \tan\left(\frac{\phi-\phi_0/2}{2}+\frac{\pi}{4}\right)\right]\ \ \text{outgoing} \label{+solution}\\
t_-\left(\phi\right)&=&\alpha\:\ln\left[ \tan\left(-\frac{\phi-\phi_0/2}{2}+\arctan\left(e^{t_0/\alpha}\right)\right)\right]\ \ \text{incoming} \label{-solution}
\eea
The left branch of the boundaries is obtained by reflecting the solutions \eqref{+solution}, \eqref{-solution} with respect to $\phi$ at $\phi_0/2$. Eqs. \eqref{+solution}, \eqref{-solution} can further be used to give an explicit relation between the parameters $\phi_0$ and $t_0$:
\bea
t_+\left(\phi_0\right)&=&t_-\left(\phi_0\right)\ \ \  \Longrightarrow \\
t_0&=&\alpha\:\ln\left[\tan\left(\frac{\phi_0}{2}+\frac{\pi}{4}\right)\right] \\
\phi_0&=&2\:\arctan\left(e^{\frac{t_0}{\alpha}}\right)-\frac{\pi}{2}
\label{t0phi0relation}
\eea
Finally, the volume of the AS is given by
\bea
V_{AS}&=&2\;\left(V_{-}-V_{+}\right) \\
V_{\pm}&=&\int_{\phi_0/2}^{\phi_0}d\phi\;\int_{0}^{t_\pm\left(\phi\right)}dt\;\sqrt{-\det\left(g\right)}=\int_{\phi_0/2}^{\phi_0}d\phi\;\int_{0}^{t_\pm\left(\phi\right)}dt\;\alpha\cosh\left(\frac{t}{\alpha}\right)\nonumber \\
&=&\alpha^2\int_{0}^{\phi_0/2}d\phi\;\sinh\left\{\ln\left[\tan\left(\pm\frac{\phi}{2}+c_\pm\right)\right]\right\} \\
c_+&=&\frac{\pi}{4},\ \ \ c_-=\arctan\left(e^\frac{t_0}{\alpha}\right)=\frac{\phi_0}{2}+\frac{\pi}{4}
\eea
\subsubsection{Results}

We sprinkle a large number of points into a region of $\ds{2}$ spacetime. Three parameters can be chosen: the scalar curvature $R$, via $\alpha$, the total number of points to be sprinkled, $N$, and the size of the region, via $t_0$ (or $\phi_0$, see Eq. \eqref{t0phi0relation}). 
We first check our sprinkling procedure by varying the size of the region and keeping the other parameters fixed. The settings are as follows:
\begin{enumerate} 
\item The total number of sprinkled points, $N$, is fixed. We use a moderate number $N=3200$.
\item The parameter $\alpha$ is set so that $R=4$, corresponding to $R_{00}=-2$. Later, the results will be compared to simulations with $R=2,6,8$.
\item 
The volume of the region is varied in order to study the effect of the density of points. We choose values
$V_\text{region}=2.0,\,1.0,\,0.5$ and $0.25$. This is achieved by combining Eq.\eqref{Aregion} with Eq.\eqref{t0phi0relation} and solving numerically for $t_0$ and $\phi_0$. 
\end{enumerate}

For each $V_\text{region}$ we do 100 sprinklings, calculate the mean of the number of sprinkled points in the region as well as the mean of points sprinkled into the AS, $N_\text{AS}$. This is compared to the theoretical prediction given by the relative size of the volume of the region and the AS, i.e.
\be
N_\text{AS}^\text{theory}=\frac{V_\text{AS}}{V_\text{region}}\:N_\text{region}
\label{NAStheory}
\ee

The proper time of the AS, $\tau$, the scalar curvature $R$ as well as the time-time component of the Ricci tensor $R_{00}$ are computed as per \cite{Roy:2012uz} i.e. from Eqs. \eqref{Rgen}-\eqref{Tgen} with $\lambda=3,\,k_1=1$. The results are summarized in Table \ref{table:NASresults}. Even though the number of sprinkled points is moderate, the agreement on $N_\text{AS}$ and $\tau$ is impressive. The curvature parameters, however, are far off the expected values. This general picture also holds for simulations with scalar curvature $R=2,6,8$.

\begin{table}[H]
	\begin{center}
	\resizebox{\textwidth}{!}{
		\begin{tabular}{c| c| c|| c| c| c|| c| c| c}
	$V_\text{region}$ & $t_0$  & $\rho$ & $N_\text{region}$ & $N_\text{AS}$ & $N_\text{AS}^\text{theory}$ & $\tau$ & $R$ & $R_{00}$\\ 
	\hline 
	2.0 & 1.321 & 1600 & $3204 \pm 5.5$ & $1234\pm 3.5$ & 1231 & 1.327 & 0.70 & $-2.1$ \\ 
	1.0 & 0.974 & 3200 & $3187 \pm 5.8$ & $1410\pm 3.9$ & 1412 & 0.978 &$1.2$ & $-2.4$ \\ 
	0.5 & 0.701 & 6400 & $3204 \pm 5.6$ & $1514\pm 3.8$ & 1513 & 0.704 & $1.5$ & $-2.9$ \\ 
	0.25 & 0.499 & 12800 & $3201 \pm 5.1$ & $1560\pm 3.4$ & 1560 & 0.500 & $1.3$ &$-3.4$  \\ 
		\end{tabular} 
		}
	\end{center}
\caption{Simulation results with sprinklings of 3200 points into a region of $\ds{2}$ with scalar curvature $R=4$. The results shown in columns 4,5,7,8 and 9 are mean values over 100 sprinklings. Error bars are standard deviations of the mean. Also shown for comparison is the expectation as given by Eq. \eqref{NAStheory}.}
\label{table:NASresults}
\end{table}

The mismatch between expectation and simulation for the scalar curvature needs some explanation. A closer look at the method used reveals that large cancellations occur when applying the result of the simulations for the distributions of $k$-chains to the defining equations \eqref{Qklam}-\eqref{R00gen}.
The obvious remedy is to use larger number of points in the AS, as well as doing more runs, thereby improving the accuracy of the individual $\av{C_k}$, $k=1,2,3$ involved. We will present results obtained along these lines below.  
Another question is whether it might be that the expansion in $\tau^2$ is slowly converging. Indeed, the situation improves a little if the proper time of the AS is small.

In the case at hand, the effect of higher order corrections on the determination of $R$ can be estimated by theoretical considerations. In $\ds{2}$ the scalar curvature and the time-time component of the Ricci Tensor are not independent, i.e.
\be
R_{00}=-\frac{1}{2}R=-\frac{1}{\alpha^2}
\label{R00Rrelation}
\ee
Using this proportionality, the quantities $Q_{k,\lambda}$ are functions of two parameters only, $\tau^2$ and $R$. Plugging in explicit expressions for the coefficients $\alpha_k, \beta_k$ for $n=2$ yields the expansion
\be
Q_{1,\lambda}=\tau^{\lambda\,n}\Bigl[1-\frac{\lambda}{48}\ R\,\tau^2\Bigr]+O\left(\tau^4\right) \label{Q1lambda_alt}
\ee

On the other hand, in the continuum we can express $Q_{1,\lambda}$ in terms of the volume of the AS, i.e. 
\bea
Q_{1,\lambda}^\text{cont}&=&\bigg(\frac{\av{C_1}}{\rho\:\zeta_1}\bigg)^\lambda \\
\zeta_1&=&\frac{1}{2},\ \av{C_1}=N_\text{AS}=\frac{V_\text{AS}}{V_\text{region}}\:N_\text{region},\ \rho=\frac{N_\text{region}}{V_\text{region}} \\
Q_{1,\lambda}^\text{cont}&=&\left(2\:V_\text{AS}\right)^\lambda \label{Q1lambda_cont}
\eea
Combining Eq.\eqref{Q1lambda_alt} with \eqref{Q1lambda_cont}, we obtain a measure for the higher order corrections in the $\tau^2$-expansion. Solving Eq. \eqref{Q1lambda_alt} for $R$ we thus have
\be
R^\text{alt}=\frac{48}{\lambda\tau^2}\Bigl[1-\frac{Q_{1,\lambda}^\text{cont}}{\tau^{\lambda\,n}}\Bigr] \label{R1_alt}
\ee  

For any given region of $\ds{2}$ with input parameters $\alpha$, $t_0$ and the number of sprinkled points, $N_\text{region}$, we can compute $R^\text{alt}$ as given in \eqref{R1_alt}. The result is as close as possible to the continuum value. Any remaining deviations must be due to the truncation of the $\tau^2$-expansion. We test this procedure for two curvature values, $R=4$ and $R=8$. We vary the size of the region considered, $V_\text{region}$, as well as the parameter $\lambda$ appearing in the definition of \eqref{Qklam}. The results are shown in Table \ref{table:R01_alt_cont}.

\begin{table}[h]
	\begin{center}
		\begin{tabular}{c| c|| c| c| c| c}
			$R$ & $V_\text{region}$ &  $\lambda$ = 3 & $\lambda$ = 1 & $\lambda$ = 0.5 & $\lambda$ = 0.1 \\ 
			\hline 
			4 & 2 & 2.89 & 3.26 & 3.36 & 3.45  \\ 
			  & 1 & 3.31 & 3.56 & 3,62 & 3.67  \\ 
			  & 0.5 & 3.61 & 3.76 & 3.79 & 3.82  \\ 
			  & 0.25& 3.79 & 3.87 & 3.89 & 3.91  \\ 
			\hline
			8 & 2 & 4.81 & 5.79 & 6.08 & 6.32  \\ 
			  & 1 & 5.78 & 6.52 & 6.73 & 6.90 \\ 
			  & 0.5 & 6.63 & 7.11 & 7.24 & 7.35  \\ 
			  & 0.25 & 7.23 & 7.51 & 7.59 & 7.64  \\
		\end{tabular} 
	\end{center}
	\caption{Estimate of the effect on $R^\text{alt}$ due to truncation of $\tau^2$-expansion.}
	\label{table:R01_alt_cont}
\end{table}

We observe that for larger regions the "continuum values" $R^\text{alt}$ are considerably off the expected input values $R=4$ or $R=8$, respectively. The situation is improved by taking regions with a smaller proper time span of the AS. Moreover, for a given fixed value of $V_\text{region}$, the agreement between Eq.\eqref{R1_alt} and the input is improved by employing smaller values of the parameter $\lambda$ in the definition of $Q_{1,\lambda}$. We therefore expect that, in simulations also, it is important to use a smaller AS in order to suppress higher order corrections. These corrections can be further tamed by resorting to smaller $\lambda$ values. 

\begin{figure}[!htp]
\begin{tabular}{cc}
\mybox{\includegraphics[width=0.385\textwidth]{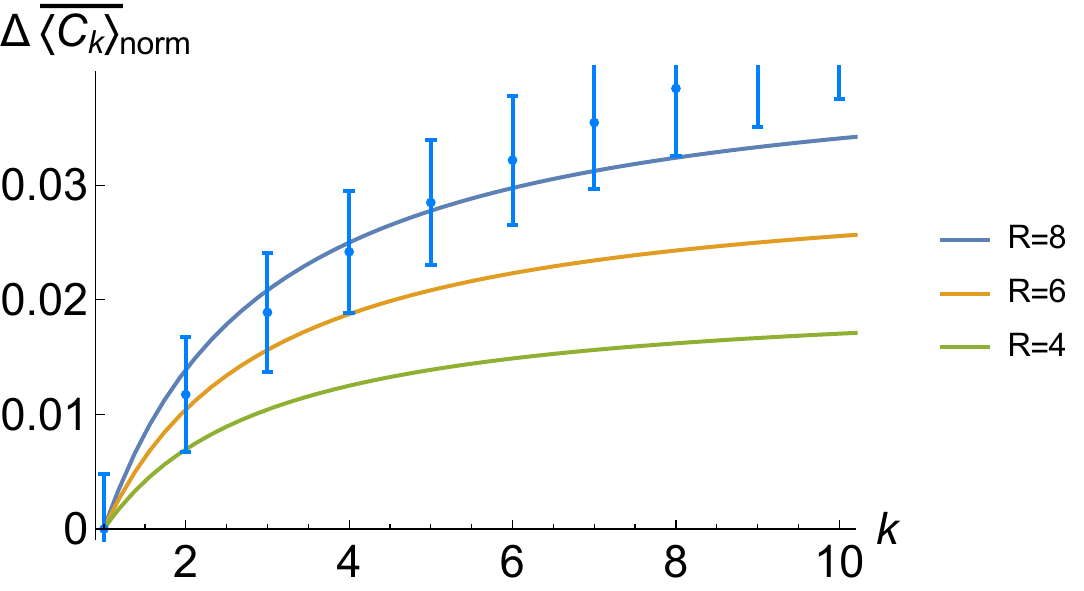}} &   \mybox{\includegraphics[width=0.385\textwidth]{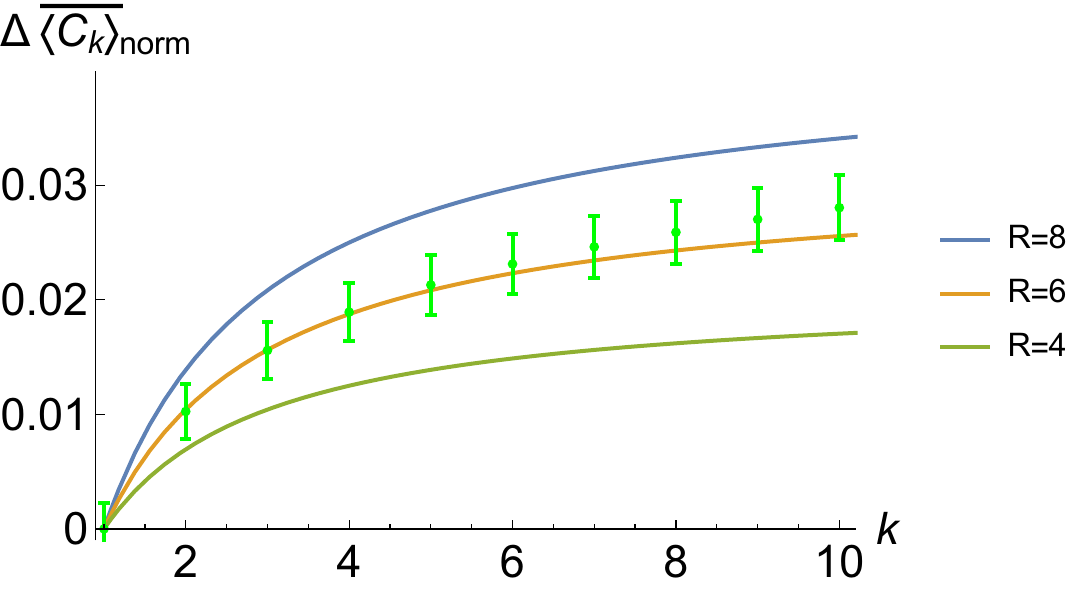}} \\
(a) $n=1600$ & (b) $n=6400$ \\[6pt]
\mybox{\includegraphics[width=0.385\textwidth]{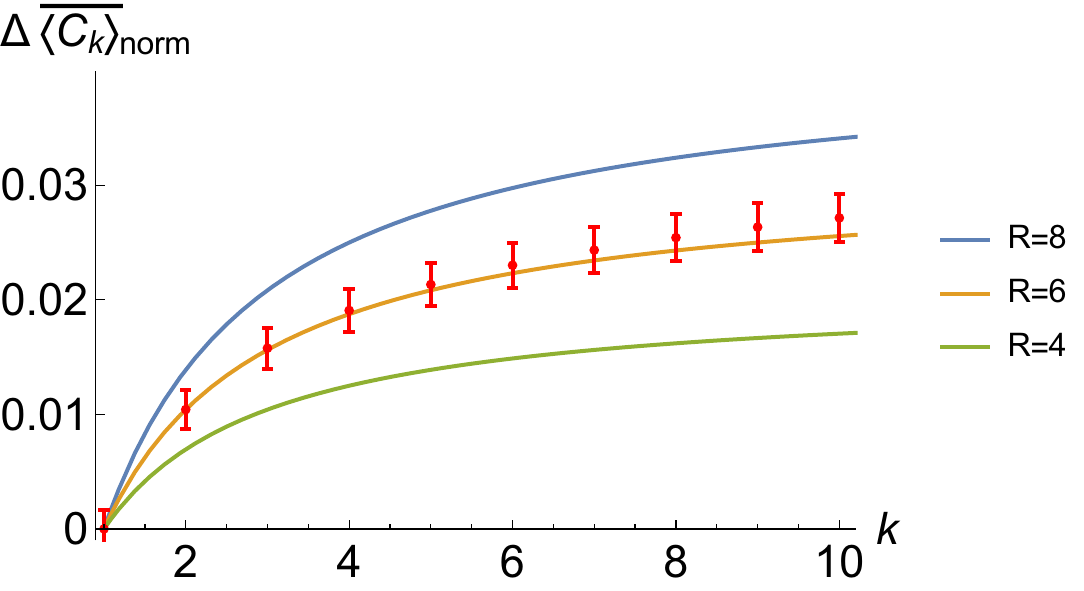}}  & 
\mybox{\includegraphics[width=0.385\textwidth]{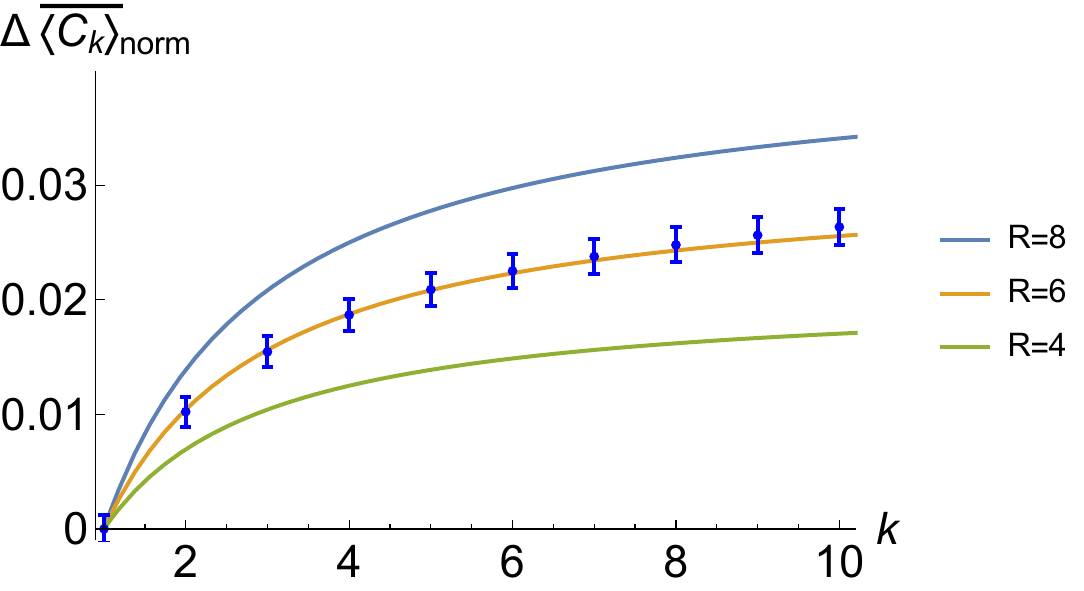}}   \\
(c) $n=12800$ & (d) $n=25600$
\end{tabular}
\caption{Relative distributions of $k$-chains normalized to $\mink{2}$, $\Delta\,\overline{\av{C_k}}_\text{norm}$, for sprinklings into $\ds{2}$ with scalar curvature $R=6$. Solid lines are theoretical expectations derived from \eqref{avgcexpansion} for $R=8,6,\text{ and}\ 4$, respectively.}
\label{fig:NNcomparison}
\end{figure}

\begin{figure}[H]
\centering
  \mybox{\includegraphics[width=7.5cm,keepaspectratio]{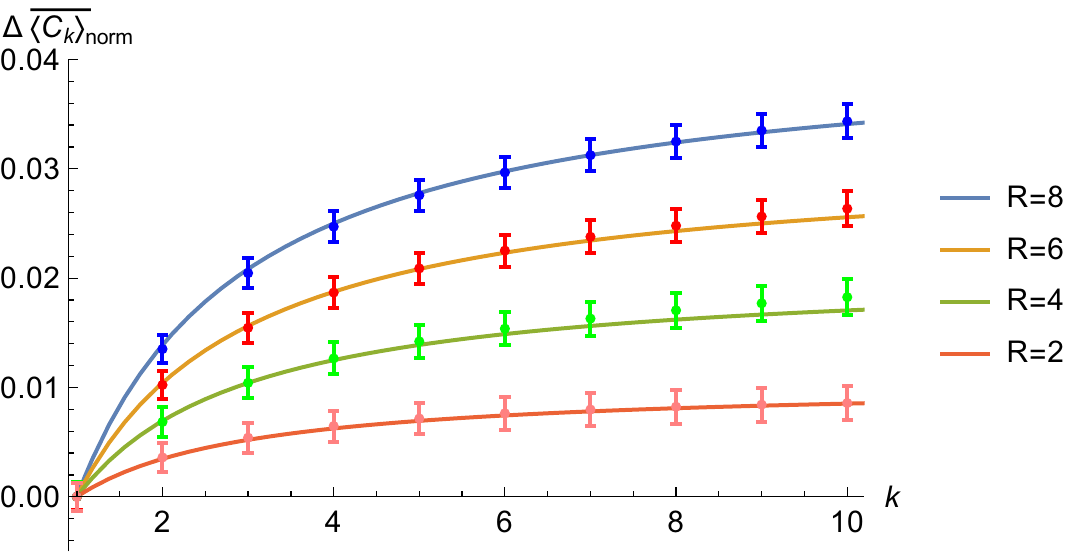}}
  \caption{The quantity $\Delta\,\overline{\av{C_k}}_\text{norm}$ from sprinklings of $N=25600$ points into $\ds{2}$ with scalar curvature $R=2,4,6$ and $8$, respectively. Solid lines are theoretical expectations.}
 \label{fig:Rcomparison}
\end{figure}

Using these insights, we present the final results for the curvature estimates of $\ds{2}$ obtained by simulations. The settings of these simulations are as follows:
\begin{itemize}
    \item The region of 
    $\ds{2}$ to be sprinkled is fixed at $V_\text{region}=0.25$. This corresponds to a proper time $\tau=0.5$ of the corresponding AS.
    \item The number of sprinkled points is increased from  $N_\text{sprinkled}=3200$ to $6400,\,12800,\,25600$.
    \item We sprinkle into $\ds{2}$ with scalar curvatures $R=2,4,6,8$. These values are input parameters.
    \item For each set of parameters, the procedure is repeated 200 times and averages are taken. Statistical error bars given are the standard deviations of the mean. 
\end{itemize}
As shown in \cite{Roy:2012uz}, the errors in the quantities of interest scale with the density as $\rho^{-1/2}$. The improvements due to large $N$ can be visualized best by looking at the deviations of the distribution of $k$-chains from the flat case. We plot $\Delta\,\overline{\av{C_k}}_\text{norm}$ against $k$ for various numbers of sprinkled points. In Fig. \ref{fig:NNcomparison} the results of the simulations are compared to the theoretical predictions \eqref{avgcexpansion} for $\ds{2}$ with $R=6$. Increasing the density of sprinkled points clearly improves the agreement between simulation and theory. For the largest density used, the data points match perfectly with the theoretical curve up to $k=6$. For $k\geq 6$, deviations are within the statistical one sigma range.
A similar plot can be used to show the results of simulations with varying scalar curvature $R$. The data in Fig. \ref{fig:Rcomparison} shows that the method is clearly able to discriminate between various $R$. Statistical errors have been brought down to a level where the differences are significant for the relative distributions of all the $k$-chains considered, $k=2,3,...,10$.  

There are several ways to determine the scalar curvature and the time-time component of the Ricci tensor from these data. One method was given in Eqs. \eqref{Rgen}, \eqref{R00gen} and the discussion around Table \ref{table:R01_alt_cont}. The results obtained with this first method are shown for a selection of parameter choices in the left half of Table \ref{table:TRresults_final}. Another possibility is to perform a least square fit to the data points of $\Delta\,\overline{\av{C_k}}_\text{norm}$ shown in Fig. \ref{fig:Rcomparison}. The advantage of this second method is that several data points are included, starting from $k=1$ up to $k_\text{max}$. As expected for method 1, agreement between simulations and expectation is improved by using a smaller parameter $\lambda$. The results also vary depending on the actual $k$-chains used, again less so for smaller $\lambda$. 
\begin{table}[!h]
	\begin{center}
	\resizebox{\textwidth}{!}{
		\begin{tabular}{c| c|| c c| c c| c| c}
			       &   & \multicolumn{4}{c|}{Method 1: Eqs. \eqref{Rgen}, \eqref{R00gen} }  & \multicolumn{2}{c}{Method 2: best fit} \\
			 input & simulation   & \multicolumn{2}{c|}{$\lambda=3$}&\multicolumn{2}{c|}{$\lambda=0.5$} &    & \\
			 \cline{3-2}\cline{4-2}
			 \cline{5-2}\cline{6-2}
			       &   & $k_1=1$ & $k_1=3$ & $k_1=1$ & $k_1=3$ & $k_\text{max}=4$ & $k_\text{max}=6$  \\
			\hline 
			$R=2$ & $R$ & 1.41& 1.97 & 1.57 & 2.09 & 2.15 & 2.27   \\
			$R_{00}=-1$ & $R_{00}$ & -1.66 & -1.22 & -1.63 & -1.21 & -0.99 & -0.94   \\
			\hline
			$R=4$ & $R$ & 3.20 & 4.12 & 3.57 & 4.37 & 3.19 & 2.96  \\
			$R_{00}=-2$ & $R_{00}$ & -1.80 & -1.25 & -1.71 & -1.21 & -2.33 & -2.43   \\
			\hline
			$R=6$ & $R$ & 4.98 & 5.42 & 5.87 & 6.05 & 5.51 & 5.26 \\
			$R_{00}=-3$ & $R_{00}$ & -2.61 & -2.36 & -2.39 & -2.28 & -3.08 & -3.19   \\
			\hline
		\end{tabular} 
		}
	\end{center}
	\caption{Results of simulations of $\ds{2}$ with $\tau=0.5$ and $N=25600$, for various input parameters $R$. The average was taken over 200 runs.}
	\label{table:TRresults_final}
\end{table}

Although these systematic errors are relatively large, the method yields qualitatively correct answers for the three curvatures considered. The results obtained by the least square fit of method 2 are even more satisfactory, except for the case $R=4$, where the scalar curvature appears to be somewhat small. However, the theoretical expectations, i.e.. $R=2,4\ \text{and}\ 6$, respectively, with $R_{00}=-R/2$, still yield a good fit to the data. The sum of squared errors (SSE), weighted with the squares of statistical errors of the data points, is always smaller than $0.06$ per degree of freedom. The quality of such fits is evident from Fig. \ref{fig:Rcomparison}, where the deviations of the data from theory are well below the one-sigma level.

\subsection{FLRW Spacetime}
\label{curvatureflrw}

We work with $n=3$ because this case needs less computing power and is easy to visualize. The generalization to $n=4$ is straightforward.  
\subsubsection{Choice of region, coordinates and parameters}
We work with coordinates $t,r,\phi$ in which the invariant line element is 
\be \label{flrwmetric}
ds^2=-dt^2+a^2(t)\bigg[\frac{dr^2}{1-kr^2}+r^2d\Omega_{n-2}^2\bigg]
\ee
Here we consider only flat spatial slices, i.e. $k=0$. The scale factor can be set to\footnote{There is a scale hidden in the definition of $a(t)$ which defines the size of the spatial sections at fixed time. We have chosen it such that $a(1)=1$.}
\be
a(t)=t^q
\ee
with possible values $q=3/2$ (matter dominated universe), $q=1/2$ (radiation dominated) or $q=1$ (vacuum dominated).

The corresponding expressions for $R$ and $R_{00}$ for $k=0$ are given by
\bea
R_{00}&=&\frac{-2\,\ddot{a}}{a}=\frac{-2 q(q-1)}{t^2}
\label{3dflrwr00}\\
R&=&2\bigg(\frac{2\,\ddot{a}}{a}+\frac{\dot{a}^2}{a^2}\bigg)=\frac{2 q(3q-2)}{t^2}
\label{3dflrwr}
\eea

\begin{figure}[!htp]
  \begin{subfigure}[b]{0.55\textwidth}
  \includegraphics[height=5cm,width=5cm]{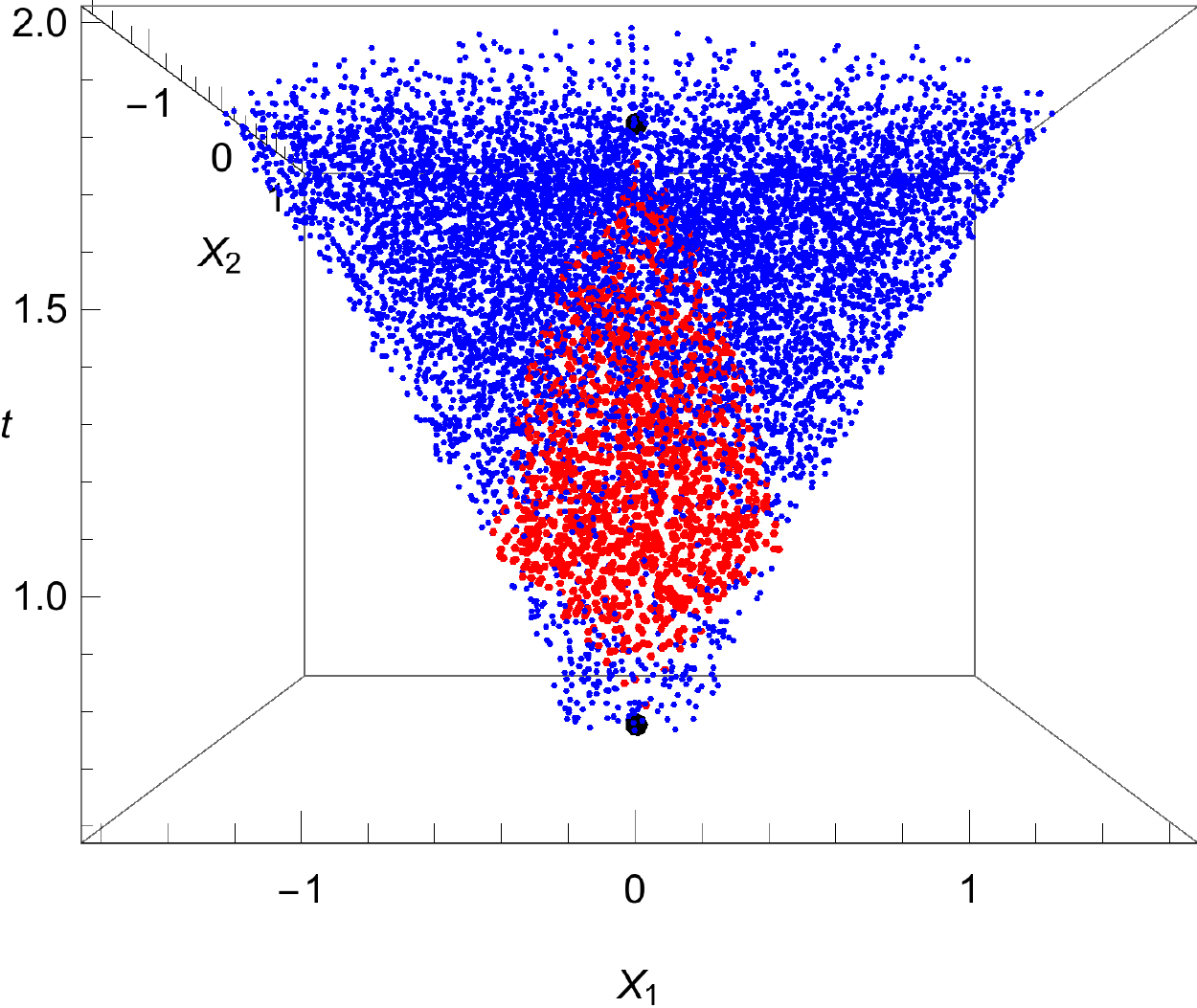}
  \caption{}
  \label{AS_points}
  \end{subfigure}
  \begin{subfigure}[b]{0.45\textwidth}
  \includegraphics[height=4.6cm,width=4.4cm]{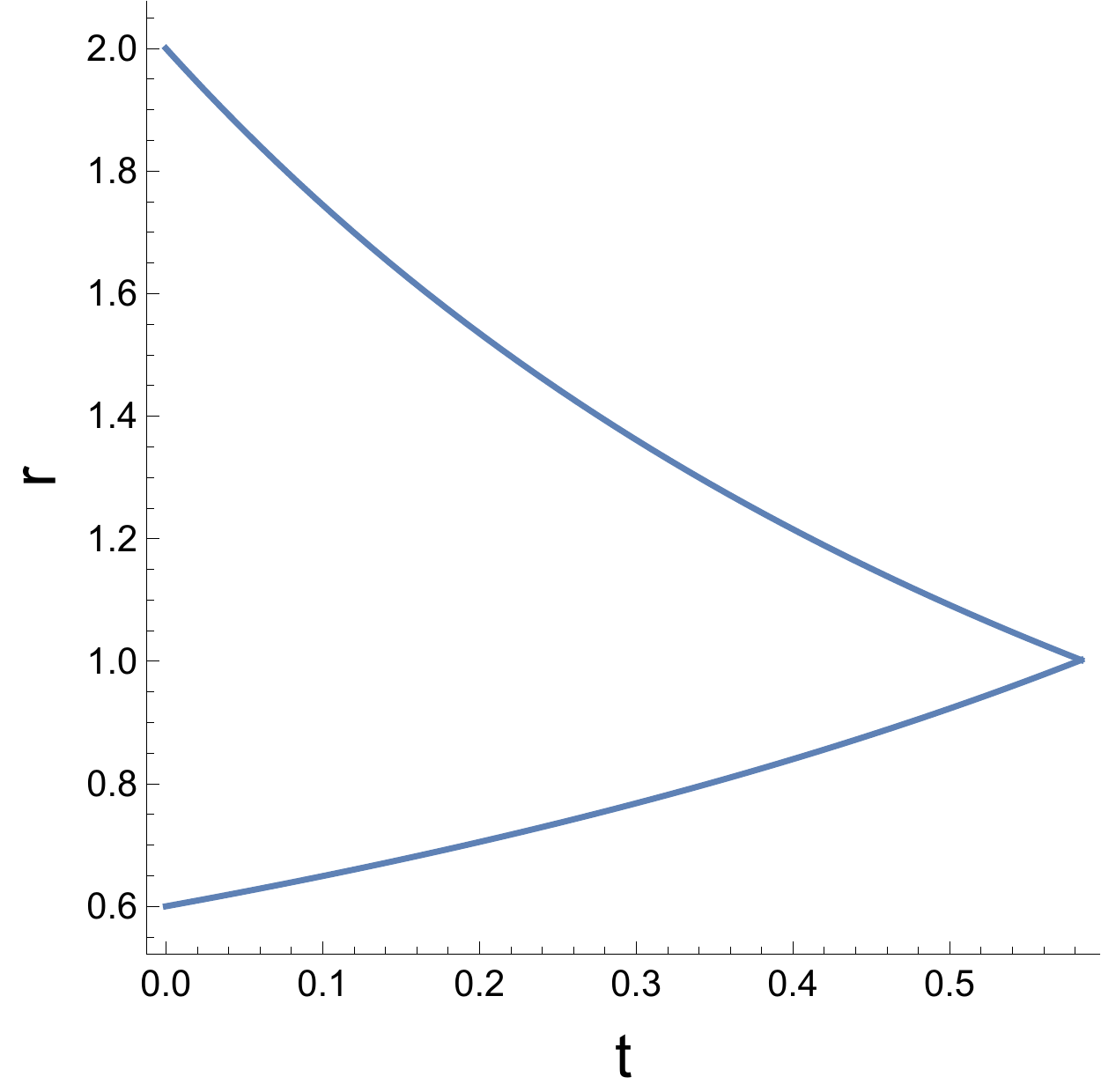}
  \caption{}
  \label{AS_example}
  \end{subfigure}
  \caption{(a) shows a causal set sprinkled in $\flrw{3}$ spacetime, a corresponding AS is shown in red and the endpoints are in black. (b) shows the shape of null lines }
  \end{figure}

In the simulations we need to sprinkle\footnote{Sprinkling details are discussed in the appendix.} into a region of $\flrw{3}$ spacetime which is well suited to construct an AS between two points $P_1, P_2$. The region can be parametrized as follows:
\bea
t_1 \leq t \leq t_2 \\
0 \leq r \leq r_1 \\
0 \leq \phi < 2\pi
\eea

\subsubsection{Creating the Alexandrov set}

Since we use only the AS for computations, it is efficient to pick the radial parameter $r_1$ such that all points of the AS between $P_1=(t_1, 0, \phi)$ and $P_2=(t_2, 0, \phi)$ are contained in the sprinkled region. To construct the AS we determine radial null rays emanating from $P_1$ in forward direction. Due to axial symmetry these null rays yield the forward light cone of $P_1$. Likewise, the backward lightcone of $P_2$ is given by the radial null rays ending on $P_2$. They are given by 
\bea
ds^2&=&0=-dt^2+a(t)^2 dr^2 \\
\frac{dr}{dt}&=&\pm \frac{1}{a(t)}=\pm t^{-q} \label{Diffeq}
\eea

Using boundary conditions $r_+(t_1)=0$, $r_-(t_2)=0$ the solution to Eq. \eqref{Diffeq} is
\bea
r_+(t)&=&\frac{1}{1-q}\left\{t^{1-q}-t_1^{1-q}\right\} \\
r_-(t)&=&\frac{1}{1-q}\left\{t_2^{1-q}-t^{1-q}\right\},\qquad q\neq 1
\eea
and 
\bea
r_+(t)&=&\ln{\frac{t}{t_1}} \\
r_-(t)&=&\ln{\frac{t_2}{t}},\qquad q=1
\eea
The time $t_m$ where the forward and the backward lightcone meet is implicitly defined through
\be
r_+(t_m)=r_-(t_m)\equiv r_1
\ee
where $r_1$ is the maximal radius of the region.

Finally we calculate the volume of the AS between $P_1$ and $P_2$
\be
V_+=\pi\int\limits_{t_1}^{t_m} \left[a(t) r_+(t)\right]^2 dt, \quad
V_-=\pi\int\limits_{t_m}^{t_2} \left[a(t) r_-(t)\right]^2 dt, \quad
V_{AS}=V_++V_-
\ee
This will be used in the determination of the density of the sprinkled causal set.    

\subsubsection{Results}

We now present the results of our simulations. The settings are as follows:
\begin{itemize}
\item We pick $q=3/2$ so that $R\,t^2=7.5$ and $R_{00}\,t^2=-1.5$.
\item The AS is fixed by initial and final points $P_1=(t_1, 0, \phi)$ and $P_2=(t_2, 0, \phi)$ with $t_1=0.6$, $t_2=1.6$ respectively. The time where the forward and the backward lightcone meet is then $t_m=0.923$, and the volume of the AS is fixed, $V_\text{AS}=0.241$.
\item The total number of points sprinkled varies between $N=12800$ and $N=102400$. This corresponds to the number of points in the the AS of $N_\text{AS}=2442$ and $N_\text{AS}=19544$, respectively. For the final results we use the highest density of points. 
\item The sprinkling procedure is repeated 100 times and averages are taken over this ensemble. The statistical error bars given are standard deviations of the mean.
\end{itemize}

\begin{figure}[!htp]
\centering
  \includegraphics[width=0.8\textwidth,keepaspectratio]{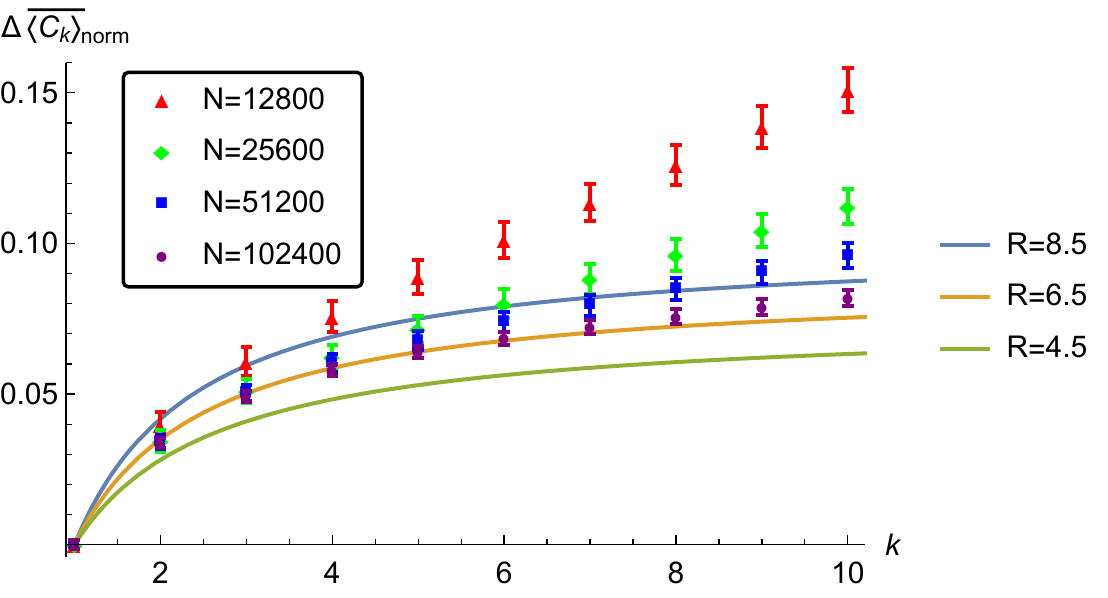}
  \caption{Relative distributions of $k$-chains normalized to $\mink{3}$, $\Delta\,\overline{\av{C_k}}_\text{norm}$, for sprinklings into $\flrw{3}$. Solid lines are theoretical expectations derived from \eqref{avgcexpansion} for scalar curvatures $R=8.5,6.5,\text{and}\ 4.5$, keeping the time component $R_{00}=-1.5$ fixed. The data points are obtained from sprinklings with varying $N$.}
 \label{fig:FLRW_NNcomparison}
 \end{figure}

Due to the dimension $n=3$, the simulations are computationally more demanding than in the previous example of $\ds{2}$ . In order to obtain reliable results, the number of sprinkled points (and hence the density of points) is increased up to $N=102400$, corresponding to approximately $20000$ points in the AS. The dependence on the number of points sprinkled is shown in Fig. \ref{fig:FLRW_NNcomparison}. As the number of points employed is increased, the data points start to converge to common values, at least up to $k=6$. Statistical errors are acceptable for discriminating between various sets of curvature parameters $R$, $R_{00}$. We therefore stick to the simulations with the highest density of points.

The determination of the geometrical parameters $R(0)$, $R_{00}(0)$ is nevertheless problematic. Firstly, both of these vary considerably over the AS, c.f. Eqs. \eqref{3dflrwr00}, \eqref{3dflrwr}. At best we can expect to obtain an average value of these parameters. While it would be desirable to use a smaller AS, the correspondingly smaller proper time would also diminish the signal, i.e. the deviations from the flat case in the perturbative expansion \eqref{avgcexpansion} would be too small. As mentioned earlier, cancellation of individual contributions to $R(0)$, $R_{00}(0)$ occurs here as well and prevents us from using a smaller AS. Our numerics confirm this observation - the region we choose is therefore a compromise. The difficulty in determination of $R(0)$, $R_{00}(0)$ at different time slices also reduces to picking the right region. Here we use a fixed region as a demonstration of the formalism. We also note that the determination of scale invariant quantities like $R(0)\tau^2$, $R_{00}(0)\tau^2$ (c.f. method 2 below) should in principle preclude the need for repeating the process at various time slices. However this needs to be checked rigorously.  

Secondly, there are multiple way of determining the sought parameters from the simulated data of the distributions of $k$-chains. Ideally, the different methods should yield comparable results, as was true to some extent for $\ds{2}$. In the case of $\flrw{3}$, there is less agreement in the results obtained by different methods. Table \ref{table:FLRWresults} summarizes the results according to the two methods considered.
\begin{table}[!h]
	\begin{center}
	\resizebox{\textwidth}{!}{
		\begin{tabular}{ c|| c c| c c| c c| c| c}
			       & \multicolumn{6}{c|}{Method 1: Eqs. \eqref{Rgen}, \eqref{R00gen} }  & \multicolumn{2}{c}{Method 2: best fit} \\
			 simulation   & \multicolumn{2}{c|}{$\lambda=3$}&\multicolumn{2}{c|}{$\lambda=1$}&\multicolumn{2}{c|}{$\lambda=0.5$} &    & \\
			 \cline{2-2}\cline{3-2}
			 \cline{4-2}\cline{5-2}\cline{6-2}\cline{7-2}
			 & $k_1=1$ & $k_1=2$& $k_1=1$ & $k_1=2$ & $k_1=1$ & $k_1=2$ & $k_\text{max}=4$ & $k_\text{max}=6$  \\
			\hline 
			$R$ & 3.45& 6.07 & 5.10 & 7.10 & 5.54 & 7.34 & 5.67 & 5.36   \\
			$R_{00}$ & -1.73 & -0.57 & -1.44 & -0.35 & -1.34 & -0.29 & -1.51 & -1.62   \\
			\hline
			$\tau$ & 1.001 & 0.989 & 0.998 & 0.987 & 0.997 & 0.987 & $-$ & $-$ \\
			$n$ & 2.999 & 2.986 & 2.996 & 2.985 & 2.995 & 2.984 & $-$ & $-$   \\
			\hline
		\end{tabular} 
		}
	\end{center}
	\caption{Results of simulations of a patch of $\flrw{3}$ between $t_1=0.6$ and $t_2=1.6$. The total number of points sprinkled was $N=102400$, averages were taken over 100 runs.}
	\label{table:FLRWresults}
\end{table}

As expected, the results for $R$, $R_{00}$  obtained by Method 1 vary considerably, depending on the choice of parameters $\lambda$ and $k_1$. While the scalar curvature appears to be in the ballpark of the expected value, the time-time component of the Ricci tensor comes out too small, in particular if higher $k$-chains are used. The best fit values of Method 2 are less sensitive to the inclusion of higher $k$-chains. In particular, $R_{00}$ perfectly matches the theoretically expected value. The weighted SSE of these fits is smaller than $0.05$. The proper time $\tau$, is very close to the expected value $\tau=1.0$, irrespective of the method used. For completeness, we have also listed the dimension $n$ calculated according to the formulae given in the next section. This parameter can be predicted very reliably.

\section{Dimensional Estimates}
\label{dimension}

In the examples considered so far it was assumed that the dimension of the spacetime in which the causal set embeds is known. In general, given a causal set, the dimension has to be deduced from it. A number of methods of obtaining the dimension from a causal set have been discussed previously \cite{Meyer:1988uy,Reid:2003,Roy:2012uz,Aghili:2018fkd}. We first show that for the causal sets obtained by sprinkling into $\ds{2}$ and $\flrw{3}$ spacetimes described in the previous section, the dimension can be obtained along the lines of \cite{Roy:2012uz} up to an ambiguity inherent to the dimension estimator. With a slight generalization of this method the ambiguity can be removed, with stable results for the dimension even for sets with a moderate number of sprinkled points. Moreover, by using higher $k$-chains, we obtain a hierarchy of dimension estimators valid for any causal set embedding into an RNN of a Lorentzian manifold. The examples studied here yield stable results for the dimension, at least for $k\leq 10$, the maximal length of $k$-chains considered in the numerical simulations. The question of how much of these results can be carried over to causal sets that are not manifoldlike is addressed in section \ref{non-manifold}.

The first dimension estimator for causal sets using $k$-chains was given by Myrheim and Meyer \cite{Myrheim:1978ce, Meyer:1988uy}. It was observed that in $\mathbb{M}^n$ the ratio of the distribution of 2- and 1-chains
\be
f_{2,\eta}\left(n\right)\equiv \frac{\left<C_2\right>_\eta}{\left<C_1\right>_\eta^2}=\frac{\Gamma\left(n+1\right)\,\Gamma\left(\frac{n}{2}\right)}{4\,\Gamma\left(\frac{3n}{2}\right)}
\label{f2eta}
\ee
is only a function of $n$. If a causal set represents flat spacetime, the comparison of the analogous expression for the causal set, $f_{2,CS}$, with Eq. (\ref{f2eta}) yields an estimate of $n$. The dimension is thus obtained from purely order theoretic information.

An obvious generalization is to consider the quantities 
\be
f_{k}\equiv \frac{\left<C_k\right>}{\left<C_1\right>^k}=\overline{\av{C_k}},\quad k=2,3,4\,...\ ,
\label{fkCS}
\ee
which can be calculated for any causal set. Again, if a causal set is to be approximated by $\mathbb{M}^n$, the distribution of $k$-chains should follow the predictions of Meyer
\be
f_{k,\eta}\left(n\right)\equiv \frac{\left<C_k\right>_\eta}{\left<C_1\right>_\eta^k}=\chi_k\left(n\right),\quad k=2,3,\,...\ ,
\label{fketa}
\ee
where the functions $\chi_k\left(n\right)$ are defined in eq. \eqref{chikdef}.

\begin{figure}[h!]
	\begin{center}
		\includegraphics[width=0.8\textwidth]{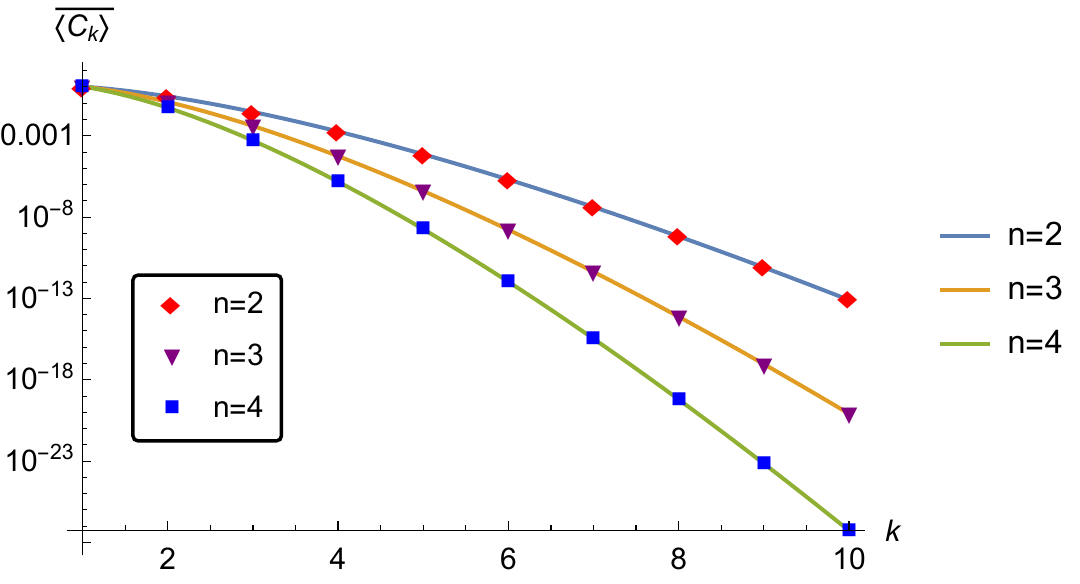}
		\caption{A log plot of the distribution of $k$-chains for $\mathbb{M}^n$ with $n=2,3,4$ dimensions. Solid lines are theoretical predictions according to Meyer \cite{Meyer:1988uy}.}
		\label{Ckbar_Minkowski}
	\end{center}
\end{figure}

The predictions of eq. \eqref{fketa} are tested by sprinkling into an AS in $\mathbb{M}^n$. For each dimension $n=2,3,4$ the procedure is repeated 100 times, and averages are taken to obtain the quantities $\left<C_k\right>$ for $k$-chains up to $k=10$. In Fig. \ref{Ckbar_Minkowski} the results of the simulations are compared to the theoretical expectations of Eq. (\ref{fketa}). Statistical errors are $\leq 1\ \%$ for $k\leq 4$ and up to $4\ \%$ for the higher $k$-chains.
 
The excellent agreement between simulations and theory can be used further. Firstly, the dimension estimator of Myrheim and Meyer is extended from one number, $f_2$, to a distribution of numbers $f_k$ over a large range of $k$-values, clearly discriminating between the dimensions $n=2,3,4$. Secondly, we may ask how much of this behaviour persists in the case of causal sets that embed into Lorentzian manifolds with curvature. Finally, since the relative distributions of $k$-chains are purely order theoretic quantities, we may even consider causal sets that do not embed into smooth manifolds. 

In the case of curved spacetimes, strictly speaking, the dimension cannot be read off directly from a plot like that in Fig. \ref{Ckbar_Minkowski}. There is no general dimension estimator known. However, in an RNN around a given spacetime point, a perturbative expansion can be used. In such cases, curvature corrections are small and the behaviour of the relative distribution of $k$-chains, $f_k$, is expected to be relatively close to the Minkowski behaviour $\chi_k\left(n\right)$. Therefore, in Fig. \ref{Ckbar_Minkowski} such spacetimes cannot be distinguished from the flat case.

\begin{figure}[!t]
	\centering
	\begin{subfigure}[b]{0.8\textwidth}
		\centering
		\includegraphics[width=\textwidth]{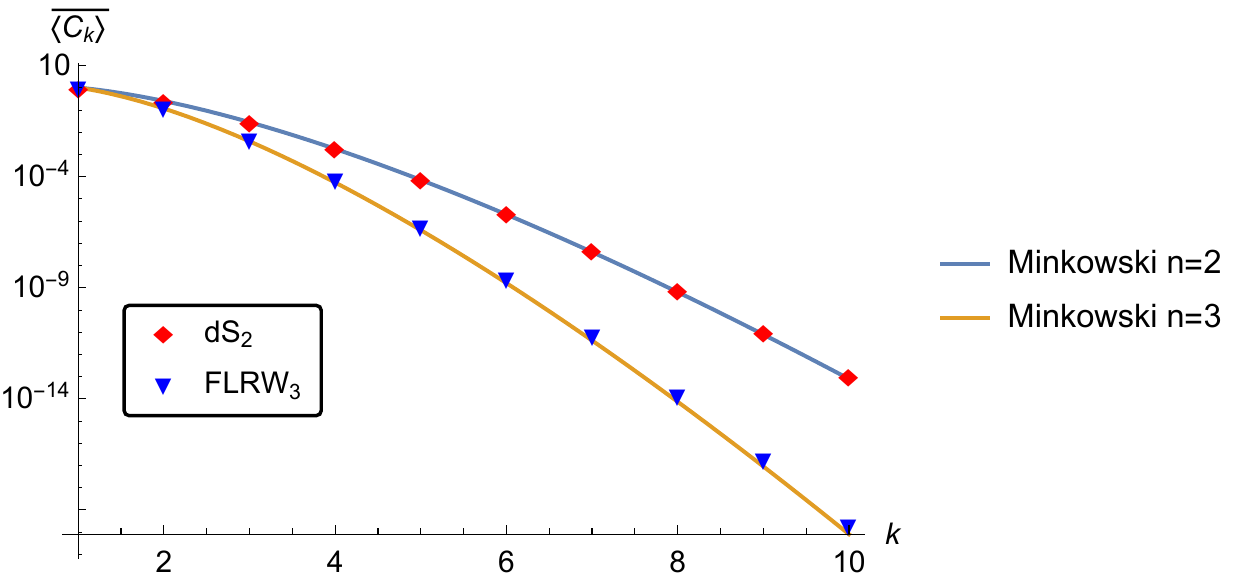}
		\caption{}
		\label{fig:LogCkbarcomp}
	\end{subfigure}
	\begin{subfigure}[b]{0.8\textwidth}
		\centering
		\includegraphics[width=\textwidth]{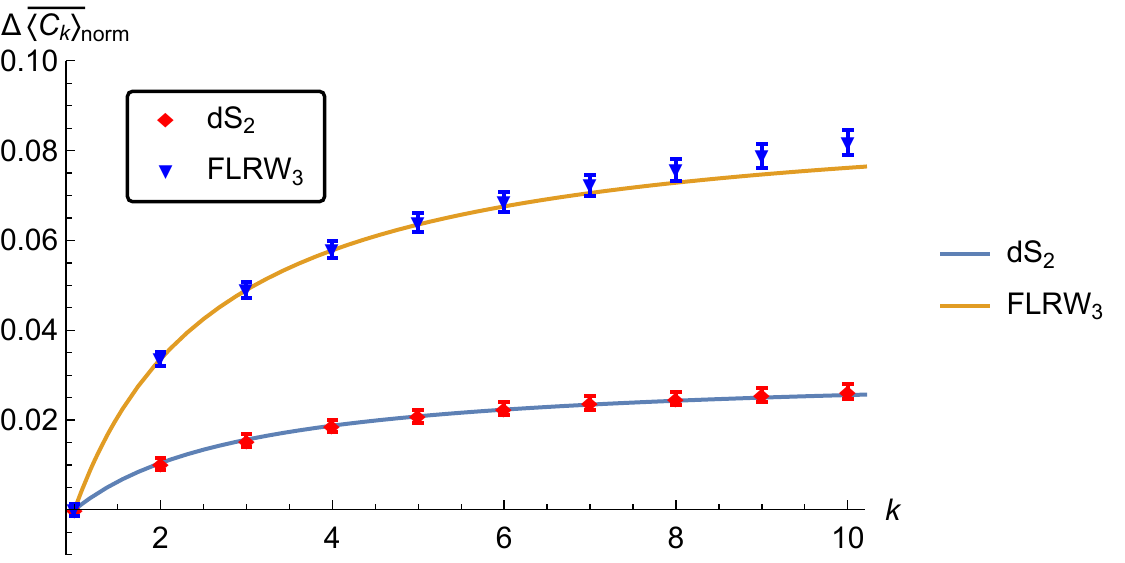}
		\caption{}
		\label{fig:LinCkbarnormcomp}
	\end{subfigure}
	\caption{Simulations of $\ds{2}$ and $\flrw{3}$ compared to theoretical expectations. Shown are (a) the log plot of $\overline{\left<C_{k}\right>}$ against $k$, where solid lines are predictions for Minkowski spacetime in $n=2$ and $n=3$ dimensions, and (b) a linear plot of $\Delta\,\overline{\av{C_k}}_\text{norm}$. Here solid lines are theoretical expectations for $\ds{2}$ with $R=6, R_{00}=-3$ and $\flrw{3}$ with $R=5.2, R_{00}=-1.86$, respectively.}
	\label{fig:Ckbarcomp}
\end{figure}

The two examples of curved spacetime considered in previous sections confirm this expectation. In Fig. \ref{fig:Ckbarcomp} the results of simulations for $\ds{2}$ and $\flrw{3}$ are compared to theoretical expectations. In the log plot of Fig. \ref{fig:Ckbarcomp} a) the relative distribution of $k$-chains shows only small deviations from the $n=2$ and $n=3$ Minkowski behaviour. The plot thus indicates the correct dimension. The deviations from the flat case are nevertheless significant, as seen in Fig. \ref{fig:Ckbarcomp} b). Here the relative distribution of $k$-chains is normalized with respect to the flat case result, c.f. Eq. \eqref{delCkbarnorm}. In flat spacetime this quantity vanishes identically by definition. Here the signal is small, but clearly visible due to small statistical errors.

A dimension estimator for an RNN in a Lorentzian manifold in terms of the distributions of $k$-chains was given in \cite{Roy:2012uz}. The quantities $\left<C_k\right>, k=1,...,4$ are combined such that the weighted sum adds up to zero, allowing us to solve for the dimension $n$. We generalize this procedure slightly in order to include higher order $k$-chains:
\bea
f_{\text{dim}}\left(n;k_1,\mu\right) &\equiv&\sum_{l=0}^3 (-1)^{l}\binom{3}{l} j_{k_1+l}(n) \left(\frac{\left<C_{k_1+l}\right>}{\chi_{k_1+l}}\right)^{\mu/(k_1+l)}=0 \label{dimest_gen}\\
j_k(n)&=&\left(kn+2\right)\left((k+1)n+2\right)
\eea
Here, $k_1\geq1$ is the smallest $k-$value for which $k$-chains are used to obtain the dimension. The parameter $\mu\geq 0$ is introduced in order to suppress higher order corrections in the perturbative expansion in $\tau^2 R$, $\tau^2 R_{00}$. Choosing $k_1=1$, $\mu=4$, we recover the expression in \cite{Roy:2012uz}.

\begin{figure}[!htp]
\centering
  \includegraphics[width=0.8\textwidth,keepaspectratio]{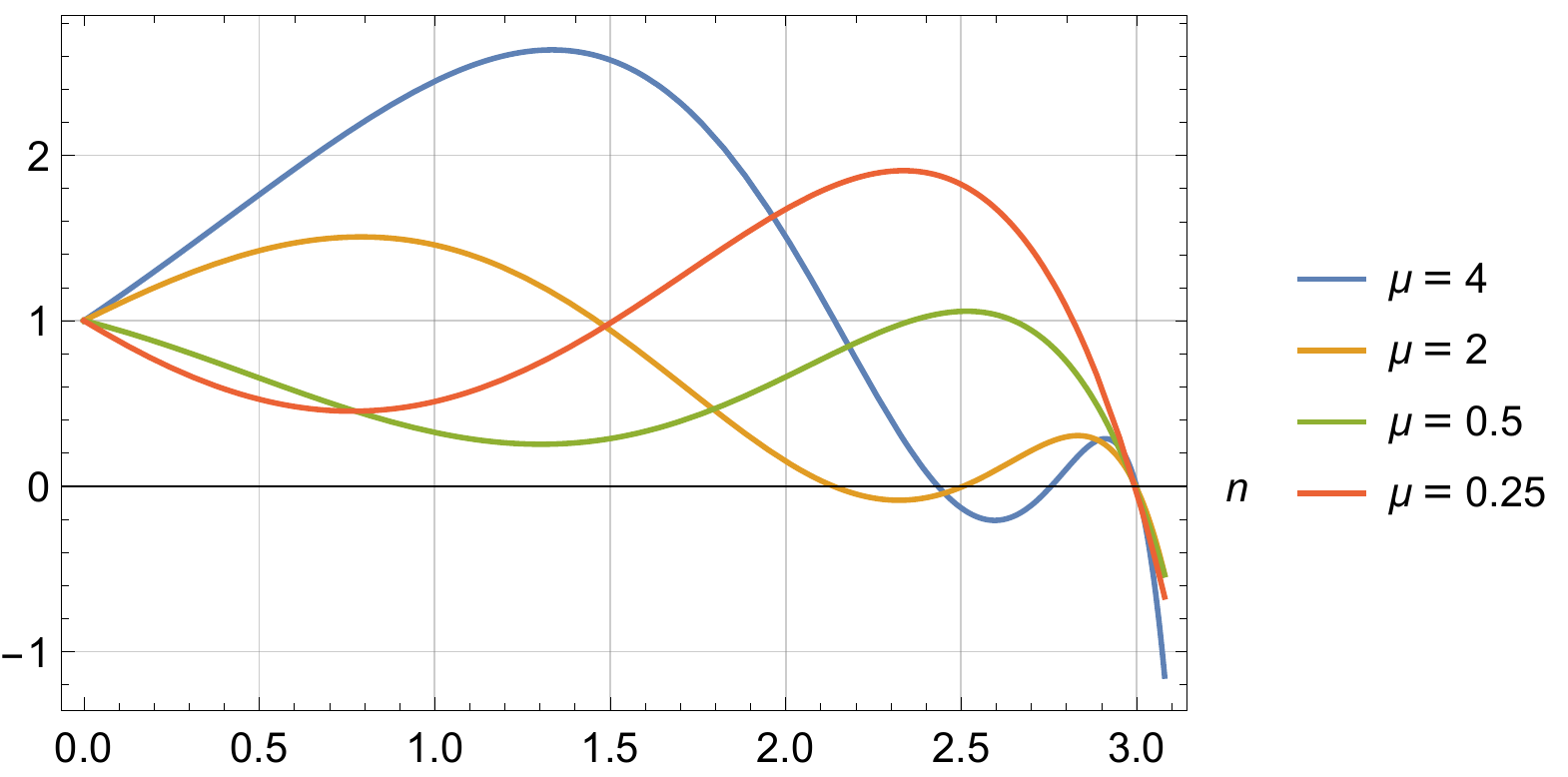}
  \caption{Normalized dimension estimator $f_{\text{dim}}\left(n;1,\mu\right)/f_{\text{dim}}\left(0;1,\mu\right)$ for various values of the parameter $\mu$. The data are taken from simulations of $\flrw{3}$ with $N=102400$.}
 \label{fig:fdim5_FRW}
 \end{figure}

We apply Eq. (\ref{dimest_gen}) to the data obtained for $\ds{2}$ and $\flrw{3}$ spacetimes.
In Fig \ref{fig:fdim5_FRW} the function $f_{\text{dim}}(n;k_1,\mu)$ is plotted against $n$ for the distribution of $k$-chains obtained from sprinkling into $\flrw{3}$ for different values of the parameter $\mu$. We observe that for values $\mu\geq2$ the function exhibits multiple zeros, which means that the dimension is not fixed unambiguously. Lowering the parameter to $\mu <1$ the ambiguity is removed and only the zero close to $n=3$ remains. 

 In Table \ref{table:nresults} the dimension $n$ is calculated from these data for various parameter values $k_1$ and $\mu$.  
\begin{table}[h]
	\begin{center}
		\begin{tabular}{c|| c c| c| c|| c| c| c}
		 & & \multicolumn{3}{c||}{$\ds{2}$}	& \multicolumn{3}{c}{$\flrw{3}$} \\
	     $k_1$ &$\mu$ & 4 & 1 & 0.25 & 4 & 1 & 0.25 \\
			\hline 
			1 & & 1.998 & 1.997 & 1.997 & 3.005 & 3.002  & 3.000 \\ 
			2 & & 1.997 & 1.997 & 1.996 & 2.988 & 2.985  & 2.985 \\
			3 & & 1.996 & 1.996 & 1.996 & 2.972 & 2.972  & 2.973 \\
		    4 & & 1.996 & 1.996 & 1.996 & 2.957 & 2.960  & 2.960 \\
		\end{tabular} 
	\end{center}
	\caption{The dimension $n$ for simulations of $\ds{2}$ and $\flrw{3}$, respectively. The parameters $k_1$ and $\mu$ are described in the text.}
	\label{table:nresults}
\end{table} 
We observe that, for both examples considered, the result is stable against variations of the minimum value $k_1$. The results are also not sensitive to the power $\mu$ in the definition of $f_{\text{dim}}\left(n;k_1,\mu\right)$. We can conclude that the dimension of causal sets that embed into regions of $\ds{2}$ or $\flrw{3}$ can be determined reliably.

\section{Examples of Non-Manifold-like Causal Sets}
\label{non-manifold}

An important issue in causal set quantum gravity is the role of non-manifold-like sets. As is well known, the great majority of causal sets do not resemble a Lorentzian manifold; instead, in the limit of a large number of elements, they are dominated by Kleitmann-Rothschild orders \cite{Kleitmann1975}.  If causal sets are to play a fundamental role in the formulation of quantum gravity, there must be a process which suppresses the contributions from these sets to the causal set path integral. In \cite{Loomis2017} it was shown that contributions from a large class of non-manifold-like sets, the so called two-level orders, are strongly suppressed in the causal set version of the Lorentzian Einstein-Hilbert action \cite{Benincasa2011}. 

Here, we follow an approach that is similar to the one proposed in \cite{Aghili:2018fkd}. If a causal set features a distribution of $k$-chains not consistent with the predictions of the RNN expansion, it indicates that the set does not embed into a Lorentzian manifold. We therefore investigate examples of non-manifold-like causal sets, apply the RNN expansion and point out differences with respect to the expected behaviour of a Lorentzian manifold. Since the RNN expansion is valid only for a small neighbourhood of some point, special attention has to be given to the proper time or size of the AS considered. In a manifold, dividing the Volume of the AS by a given factor $\kappa$ will reduce the proper time $\propto\kappa^{1/n}$, thereby reducing the expansion parameters of the RNN expansion.  

For a general causal set we will use a notion of relative size by the following procedure: Given an AS with a finite number of elements, $I[p,q]$, the proper time can be determined, up to some arbitrary scale, by finding the chain of maximal length between $p$ and $q$ \cite{Brightwell:1990ha}. Now if we randomly pick any element $q_1\in I[p,q]$ and construct a new AS with $q_1$ as final element, $I[p,q_1]$, this new AS will have a smaller proper time than the original one. If the original causal set consists of a large number of elements, the procedure can be repeated until the size is reduced by a suitable factor. The RNN formalism can then be applied to the smaller AS and tested for self consistency. In order to test these ideas, we study two models of causal sets.

\subsection{Coupled Chains Model (CCM)}

\begin{figure}[!h]
\centering
  \includegraphics[width=5cm,keepaspectratio]{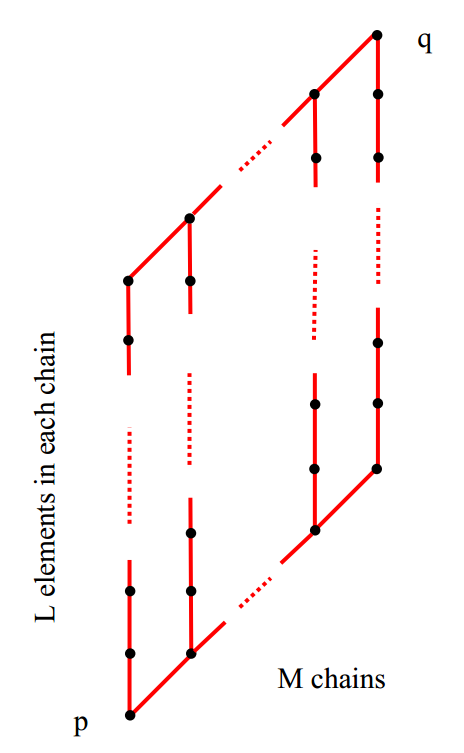}
  \caption{The causal structure of the Coupled Chains Model. Black dots denote elements, red lines causal connections.}
 \label{fig:LxMchainmodel}
\end{figure}

The model consists of an ensemble of M chains, each containing L elements that are causally connected as shown\footnote{This figure is not a spacetime diagram, it merely shows the causal connection between points.} in Fig.\ref{fig:LxMchainmodel}. In the M-direction only the lower- and uppermost elements are connected.
Choosing the causal connection this way we define the causal matrix and hence the causal set. All elements together form an AS $I[p,q]$.

\begin{figure}[!h]
\centering
  \includegraphics[width=0.8\textwidth, keepaspectratio]{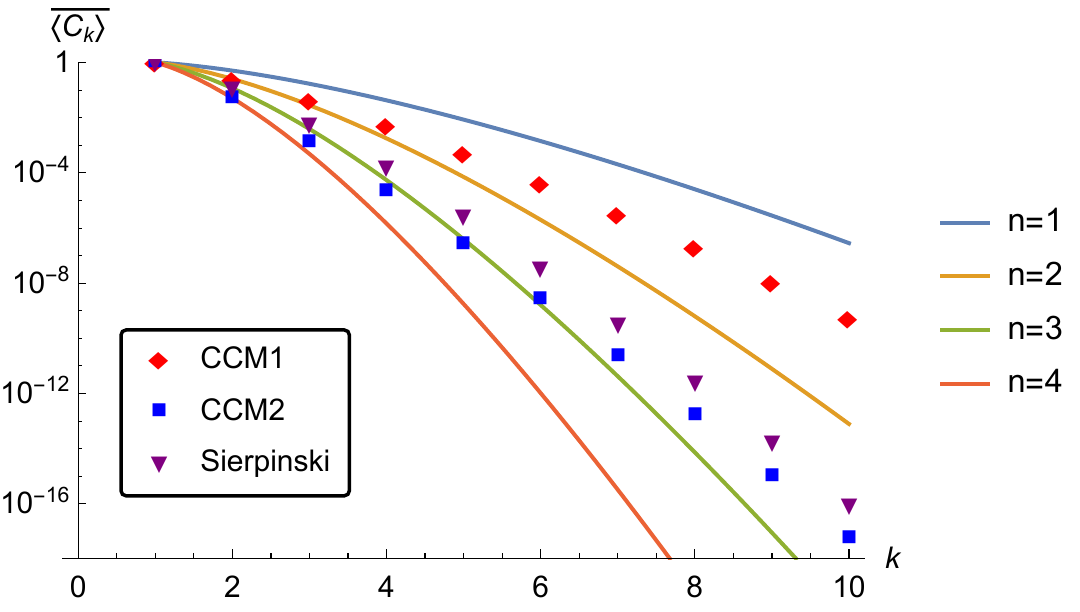}
  \caption{Relative distributions of $k$-chains for two Coupled Chains Models and the model based on a Sierpinski mesh (data points), compared to Minkowski spacetimes $\mink{n}$ (solid lines).}
 \label{fig:LxMchain_Ckbar}
\end{figure}

The distribution of $k$-chains of the CCM can be obtained by elementary combinatorics
\bea
C_k\left(L,M\right)&=&M\binom{L+M-1}{k}-\left(M-1\right)\binom{M}{k}\\
k&=&1,...,L+M-1 \nonumber
\eea
The numbers $C_k(L,M)$ characterize the causal set formed by the CCM. In Fig. \ref{fig:LxMchain_Ckbar} we compare the relative distributions of $k$-chains Eq. \eqref{fkCS}, with the predictions of Meyer for the Minkowski case with $n=1,2,...,4$. Two examples are shown: One model (CCM1) with $M\ll L$, i.e. $L=1000, M=2$, and a second model (CCM2) with $L=20, M=800$. The first model is chosen such that the Myrheim-Meyer dimension $d_\text{MM}=2$, for the second we have $d_\text{MM}\approx 4$. 

If only the Myrheim-Meyer dimension was given, the dimension of the models could not be distinguished from $\mink{n}, n=2,\,4$. However, as shown in Fig. \ref{fig:LxMchain_Ckbar}, the distributions of $k$-chains strongly deviate from Minkowski behaviour for higher $k$-values: the models cannot represent flat, n-dimensional manifolds. \footnote{The same conclusion applies to 2- or 3-level orders, like e.g. the Kleitmann-Rothschild orders.} This observation is generic and holds for a wide range of possible L and M values.

We may go one step further and compare the model to local regions of curved manifolds. We treat the numbers $C_k(L,M)$ as if they originated from averaging over an ensemble of sprinklings into a AS in a manifold $(\mathcal{M},g)$, i.e. $C_k(L,M)=\av{C_k}$. We then compute dimension, proper time and curvature according to the formulae of the RNN expansion. Inspection of the defining formulae \eqref{Rgen}-\eqref{Tgen} reveals that all quantities depend on the dimension $n$. For the approach to be self consistent, the dimension has to be fixed first from Eq.\eqref{dimest_gen}. For definiteness we fix one of the free parameters to $\mu=0.5$, but let the second vary in the range $k_1\in\{1,3,5\}$. The results are summarized in the first row of table \ref{table:CCMresults}. Not surprisingly, this procedure yields no consistent value for the dimension of the CCM. This strongly contrasts with our findings for the simulations of $\ds{2}$ and $\flrw{3}$ of section \ref{dimension}, where we found particularly stable results for the dimensions, c.f Table \ref{table:nresults}.

\begin{table}[!ht]
	\begin{center}
		\begin{tabular}{c|| c c| c| c|| c| c| c || c| c| c}
		 & & \multicolumn{3}{c||}{CCM1}	& \multicolumn{3}{c||}{CCM2} & \multicolumn{3}{c}{Sierpinski} \\
	      &$k_1$ & 1 & 3 & 5 & 1 & 3 & 5 & 1 & 3 & 5\\
			\hline 
			$n$ & & 1.73 & 1.20 & 1.08 & 2.67 & 1.66 & 1.35 & 2.11 & 1.68 &  1.41\\ 
			$\tau$ & & 1.94 & 0.96 & 0.67 & 1.71 & 0.64  & 0.30 & 1.13 & 0.58 & 0.28 \\
		    $R\,\tau^2$ & & -95.4 & -96.0 & -82.8 & -221 & -353  & -428 & -139 & -172 & -206  \\
		    $R_{00}\, \tau^2 $ & & -43.2 & -26.2 & -12.0 & -56.2 & -52.6  & -34.6 & -13.2 & 26.7 & 74.6
		\end{tabular} 
	\end{center}
	\caption{Dependence on $k$-chains used for simulated quantities of Coupled Chains Models CCM1 and CCM2, respectively. Also shown are results for the model based on a Sierpinski mesh, see subsection \ref{subsection_Sierpinski}.}
	\label{table:CCMresults}
\end{table}

Proceeding now to the calculation of $\tau$, $R$ and $R_{00}$, we first need to introduce a scale. Usually this is done by choosing an arbitrary volume and hence density in the definition of $Q_{k,\lambda}$ in \eqref{Rgen}. While the individual quantities $\tau$, $R$ and $R_{00}$ depend on the scale, the products $\tau^2 R$ and $\tau^2 R_{00}$ do not. Choosing $V_\text{AS}=1$ and $\lambda=1.0$ in Eq. \eqref{Rgen} we obtain the results given in row 2-4 of Table \ref{table:nresults}.

All quantities strongly vary with the parameter $k_1$, a consequence of the strong $n$-dependence involved in the method. These large and unstable results clearly signal the breakdown of the RNN expansion. We still have to show whether this breakdown is genuinely due to some non-manifold nature of the underlying causal set, or perhaps only due to the use of a large AS. Applying the procedure to reduce the size of the AS as discussed above, we quickly end up with an AS consisting of one single chain of causally connected elements. The RNN expansion then yields dimension $n=1$ and curvature values close to zero. The distribution of $k$-chains in Fig.\ref{fig:LxMchain_Ckbar} changes dramatically to the curve labeled with $n=1$. This transition is due to the peculiar form of the CCM models and indicates that these causal sets do not correspond to a manifold.

\subsection{Causal set based on Sierpinski triangles}
\label{subsection_Sierpinski}

\begin{figure}[!ht]
\begin{tabular}{cc}
\mybox{\includegraphics[width=0.385\textwidth]{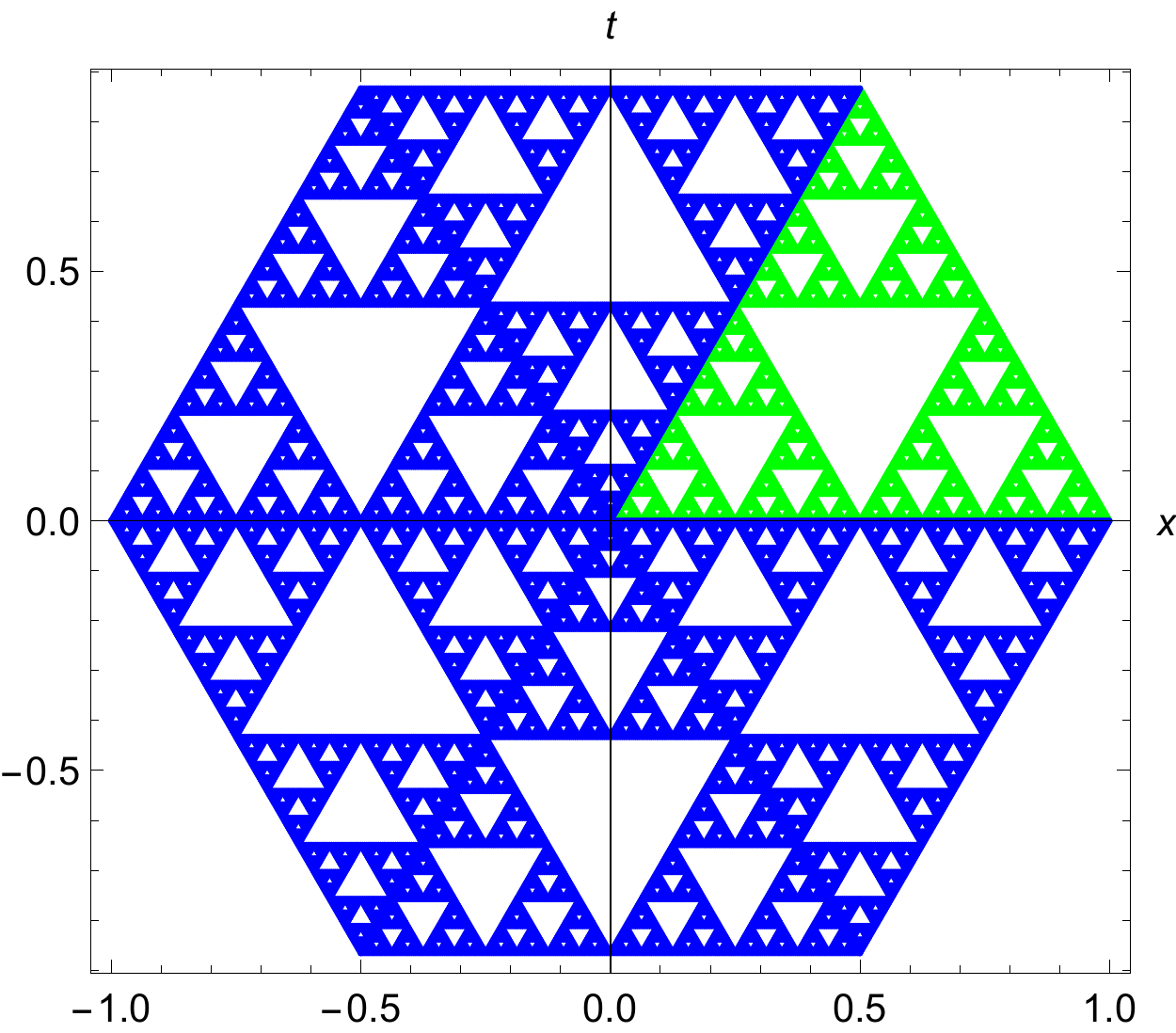}} &   \mybox{\includegraphics[width=0.385\textwidth]{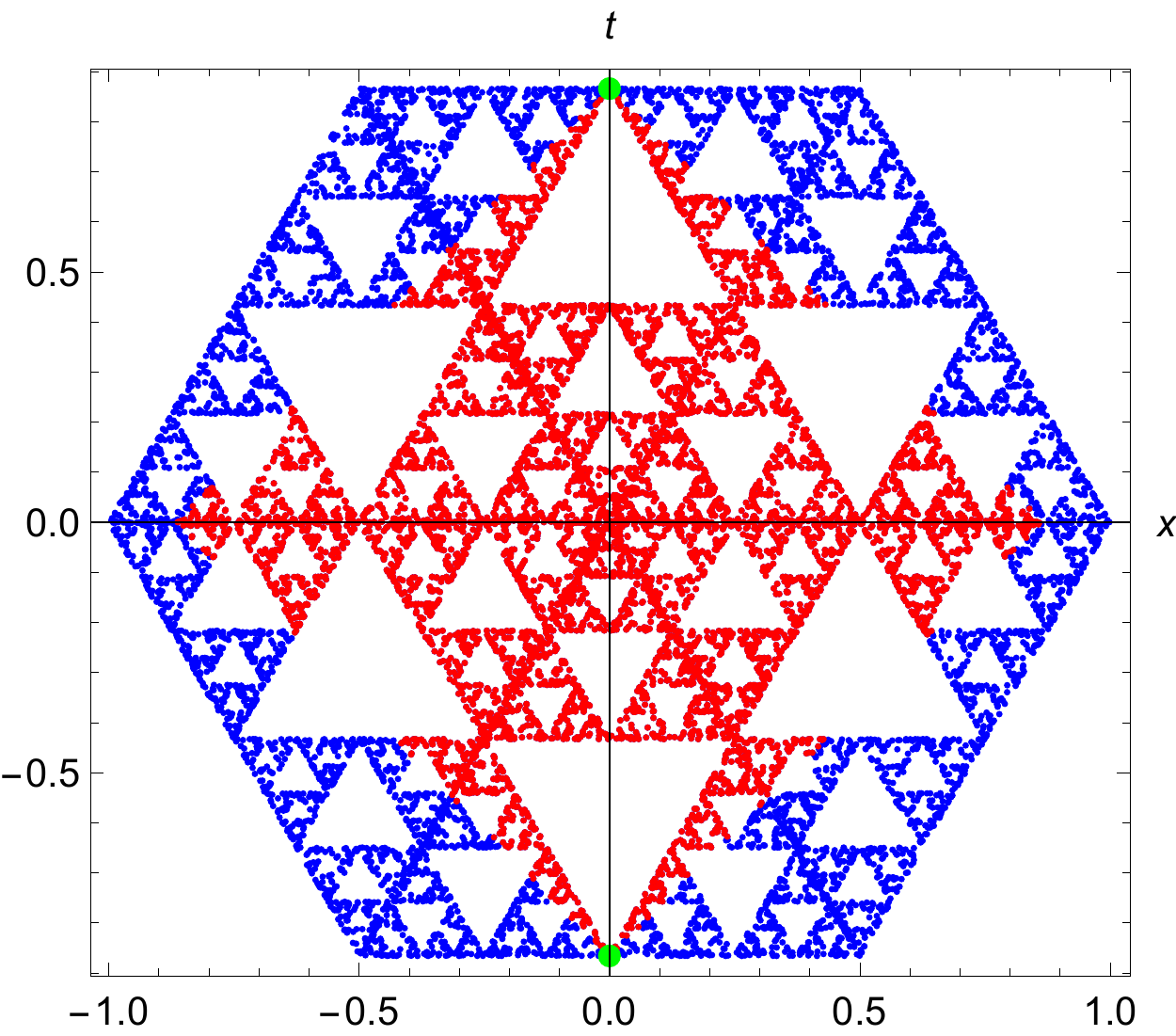}} \\
(a) Hexagonal region & (b) Sprinkling and AS \\
\end{tabular}
\caption{Construction of an AS based on Sierpinski triangles. Shown are a) Construction of hexagonal shaped region, and b) Sprinkling into this region with an AS (red points) between initial and final elements (green points).}
\label{fig:Sierpinski_mesh}
\end{figure}

The purpose of this model is to have an example resembling a patch of a continuous manifold, yet differing significantly in some aspects. We generate a causal set from a Sierpinski triangle as follows: A Sierpinski mesh of a fixed order is generated, leading to the typical picture of a Sierpinski triangle. In order to have a more regular tiling of the plane, six copies of this mesh are glued together to yield the hexagon shaped mesh of Fig. \ref{fig:Sierpinski_mesh} a). The vertices of this mesh are used as cells to which points are sprinkled by a random process. The causal relations between these points are defined to be given according to the Minkowski metric of the underlying $(x,t)$-plane. Choosing initial and final points $p,q$, the AS is given by the intersection of the forward and backward lightcones of these points, respectively. An example of this procedure is shown in Fig. \ref{fig:Sierpinski_mesh} b). Finally, the causal matrix of the AS is determined according to the metric of $\mink{2}$.

The causal set so constructed is not meant to have any physical meaning. However, given only the causal matrix as defined above, we may ask whether an ensemble of such sets could possibly embed into a Lorentzian manifold. The distribution of $k$-chains shows that this is not the case. Using Sierpinski triangles of order 14, the hexagonal region contains approximately $4\times 10^7$ vertices. Sprinkling with a probability of $0.5\times 10^{-3}$, the causal matrix has a dimension of $N_\text{AS}\approx 11300$. Repeating the procedure 100 times, the ensemble average yields a distribution of $k$-chains shown in Fig. \ref{fig:LxMchain_Ckbar}. As for the CCM the data points interpolate between integer dimensions, here between $n=2$ and $n=3$. Also, taking into account $k$-chains between $1\leq k \leq 8$, it is impossible to assign a definite dimension to this data. In Table \ref{table:CCMresults} the results obtained from the dimension estimator \eqref{dimest_gen} are seen to vary strongly with the parameter $k_1$. The same is true for the other dimensionless quantities shown in the second and third row of the table. The RNN expansion clearly fails to give sensible results. An attempt to reduce the size of the AS, thereby improving the convergence of the RNN expansion, does not work here due to the self similarity of the underlying Sierpinski mesh. Taking initial and final points of the AS in Fig. \ref{fig:Sierpinski_mesh} b) only halfway from the origin, one recovers the shape obtained from Sierpinski triangles of one order less, rescaled by a factor of two. The scale invariant quantities $n$, $R\tau^2$ and $R_{00}\tau^2$ must remain the same, except for fluctuations due to the sprinkling process. Simulations with a Sierpinski mesh of order 14 divided consecutively by factors of two confirms this expectation. The distribution of $k$-chains cannot be reconciled with the RNN expansion.

\section{Conclusions and Outlook}
\label{conclusions}

In this work we applied the formalism presented in \cite{Roy:2012uz} to causal sets obtained from a small AS in $\ds{2}$ and $\flrw{3}$ spacetimes. We showed that it is robust in determining the proper time, the scalar curvature $R$, the time-time component of the Ricci tensor $R_{00}$ and the spacetime dimension. We found that the most stable results are obtained when determining scale invariant quantities like $R\tau^2$, $R_{00}\tau^2$ and the dimension. This is due to the perturbative nature of the formalism and the use of RNNs. The correction terms in the expansion tend to be sensitive to the choice of region and they might have significant fluctuations which can be kept in check by using scale independent quantities.

From a theoretical perspective, we point out the utility of using higher chains. As we have shown, for all quantities of interest, the formalism yields an entire hierarchy of expressions implying strong consistency checks on the distribution of $k$-chains. 
We also applied the formalism to pathological causal sets constructed from gluing together sets of chains and from a Sierpinski mesh. We found that the dimension estimator does not give a stable value which then renders the estimation of other quantities meaningless as shown in Table \ref{table:CCMresults}. This suggests that these causal sets do not have any corresponding manifold structure. We conjecture this pattern to be more general - if a causal set features a distribution of $k$-chains not consistent with the RNN-expansion, it cannot be represented by an RNN in a Lorentzian manifold. We expect this criterion to be useful in distinguishing manifoldlike causal sets. 

It is worth pointing out that another test of non-manifoldlikeness based on the notion of locality on a causal set has been proposed previously \cite{Glaser:2013pca}. Instead of using chains, this test is based on a class of objects called order-intervals. While chains capture the global nature of causal relations between points, order-intervals "layer" the past light cone of a given point and are more suited in discussions of locality. It would be of interest to check if our results are compatible with this work.

Another interesting direction would be to use the scalar curvature obtained through this formalism to define a causal set action based on chains. This could be an interesting alternative to the Benincasa-Dowker-Glaser action which is based on order intervals \cite{Benincasa2011}.

Our work is a demonstration of the proof of concept and is far from comprehensive. As discussed in the introduction, the theme of reconstructing the manifold from an underlying discrete substructure runs across areas of mathematics and theoretical physics. We have restricted ourselves to the context of causal sets which are relevant to quantum gravity. Steps in this direction are pieces of an idea that all modern theories of quantum gravity point to - a discrete quantum spacetime. \\\\
\textbf{Acknowledgements}: We thank the Perimeter Institute for Theoretical Physics where this research was initiated. In particular, JK would like to thank Lee Smolin for hospitality. We also thank Sumati Surya for useful discussions. JK would like to thank the canton of Zug for the grant of a Weiterbildungsurlaub, and Paul A. Truttmann for the collaboration at an early stage of this work.

\appendix
\section*{Appendix: Sprinkling}

In the introduction we commented on why it is often convenient to work with sprinkled causal sets. Here we discuss the sprinkling procedure. \\
The procedure is an algorithm to pick $n$ points randomly from a given region of some spacetime $(\mathcal{M},g)$ at constant density $\rho$ such that eq.\eqref{numbervol} holds with $N=\av{n}$. The value of $N$ is fixed by the volume of the region and $n$ is Poisson distributed around this number (see eq.\eqref{poisson}). Repeating this process then gives an ensemble of causal sets.  

Sprinkling procedures for Minkowski and deSitter spacetimes have been discussed elsewhere \cite{Johnston:2010su,Surya:2018byh}. Here we discuss 2 methods of sprinkling in a region of $\flrw{3}$ spacetime. \\

The first method is based on the use of a global probability distribution that reflects the variation of spacial volume with time through the factor $a(t)$. The conformal coordinate system is the most convenient choice for this method and it can be obtained from eq.\eqref{flrwmetric} using the transformation\footnote{We restrict to the example used in the paper with $k=0$ and $a(t)=t^q$.} $d\eta=\displaystyle \int \dfrac{dt}{t^q}$
\be
ds^2= \Tilde{a}^2(\eta)[-d\eta^2+d\Omega_{n-1}]
\ee
with $\Tilde{a}(\eta)=((1-q)\,\eta)^{q/(1-q)}$.

To sprinkle, we first pick points randomly on the spatial part, i.e., the sphere $S^{n-1}$. One way (by no means unique) to do this is to generate normalized $n-1$-dimensional vectors. These represent points in $S^{n-1}$. The corresponding spherical coordinates can be obtained by using the standard Cartesian to spherical coordinate transformation.

Next we need the temporal part of the coordinates. The distribution of these points depends on the conformal factor. This effect can be incorporated by defining a normalised probability distribution with a probability density function equal to $\Tilde{a}^n(\eta)$ in the region of interest. Picking points from this distribution will then give us the temporal part. Combining the coordinates from the two steps, we have the required sprinkling.

The second method relies on a local probability distribution and the division of the spacetime region into cells. The choice of coordinates doesn't play an important role here and this method is easier to generalize to other spacetimes.
To implement this we need to place a point in a cell in a way that respects eq. \eqref{numbervol}. We first define the probability that the $\text{i}^{\text{th}}$ cell\footnote{Each cell is in fact labelled by 3 indices representing discretization in 3 directions, here we use only one for simplicity.} contains a point by 
\be
p_i=n\,\frac{dV_i}{V}
\ee
where $n$ is the total number of points, $V$ is the volume of the region and $dV_i$ is the volume of the $\text{i}^{\text{th}}$ cell which can be obtained from the discretized version of the volume element of $\flrw{3}$. We then generate a random number $r_i$ between 0 and 1 corresponding to each cell i. If $r_i<p_i$, we assign a point to the cell, otherwise we don't. Since $p_i$ reflects the volume of a cell, this algorithm ensures that the number-volume relation holds.

\bibliography{refs_main}{}
\bibliographystyle{ieeetr}
\end{document}